\documentclass[preprint,12pt]{elsarticle}

\usepackage[margin=1in]{geometry}
\usepackage{amssymb}
\usepackage{amsmath}
\usepackage{fixltx2e} 		
\usepackage{graphicx}
\usepackage{lineno} 			
\usepackage[normalem]{ulem}
\usepackage[title]{appendix}
\usepackage{textcomp}

\usepackage{tocloft}
\renewcommand*\cftfigpresnum{Figure~}
\usepackage{float}

\usepackage[hidelinks]{hyperref}
\makeatletter
\providecommand{\doi}[1]{%
  \begingroup
    \let\bibinfo\@secondoftwo
    \urlstyle{rm}%
    \href{http://dx.doi.org/#1}{%
      doi:\discretionary{}{}{}%
      \nolinkurl{#1}%
    }%
  \endgroup
}
\makeatother

\journal{Journal of Power Sources}

\begin{document}
\begin{frontmatter}

\title{Lithium-ion battery thermal-electrochemical model-based state estimation using orthogonal collocation and a modified extended Kalman filter}

\author[label1]{A.M.~Bizeray}
\ead{adrien.bizeray@eng.ox.ac.uk}
\author[label1]{S.~Zhao}
\ead{shi.zhao@eng.ox.ac.uk}
\author[label1]{S.R.~Duncan}
\ead{stephen.duncan@eng.ox.ac.uk}

\author[label1]{D.A.~Howey\corref{cor1}\fnref{fn2}}
\ead{david.howey@eng.ox.ac.uk}
\ead[url]{http://epg.eng.ox.ac.uk/users/david-howey}
\cortext[cor1]{Corresponding author. Tel.: +44~1865~283~476}

\address[label1]{Department~of~Engineering~Science, University~of~Oxford, Parks~Road, Oxford, OX1~3PJ, United~Kingdom}

\begin{abstract}
This paper investigates the state estimation of a high-fidelity spatially resolved thermal-electrochemical lithium-ion battery model commonly referred to as the pseudo two-dimensional model. 
The partial-differential algebraic equations (PDAEs) constituting the model are spatially discretised using Chebyshev orthogonal collocation enabling fast and accurate simulations up to high C-rates.
This implementation of the pseudo-2D model is then used in combination with an extended Kalman filter algorithm for differential-algebraic equations to estimate the states of the model. The state estimation algorithm is able to rapidly recover the model states from current, voltage and temperature measurements. Results show that the error on the state estimate falls below 1~\% in less than 200~s despite a 30~\% error on battery initial state-of-charge and additive measurement noise with 10~mV and 0.5~K standard deviations.
\end{abstract}


\end{frontmatter}



\section{Introduction}
\label{sec:introduction}
Lithium-ion batteries are widely used in electric vehicles and hybrid electric vehicles due their high energy and power density compared to other battery chemistries, and are increasingly of interest in grid and off-grid applications. However, scaling up the size of battery packs for automotive and other applications raises new safety and reliability challenges that require development of novel sophisticated battery management systems (BMSs). A BMS consists of hardware and embedded algorithms that ensure the safe and reliable operation of a pack by monitoring cells and estimating their states, such as state-of-charge (SOC) and state-of-health (SOH) \cite{Howey2015}. In order to infer unmeasurable states from the available measurements of voltage, current and temperature, a model must be solved in the BMS. In automotive applications, the model should accurately describe behaviour under the wide range of operating conditions encountered, including high current, extreme temperatures and highly dynamic loads. In addition, diagnosis and prognosis of degradation in terms of capacity and power fade is an acute challenge.

Current BMSs typically employ low-order empirical models, such as equivalent-circuit models (ECMs), which are parametrised using time- or frequency-domain experimental data \cite{Plett2004d,Hu2012,Birkl2013} for battery state estimation and control. These models have relatively low computational demands but are only valid within the narrow operating conditions in which they have been parametrised. Because the parameters of such models have little physical significance, broadening their validity range requires a large amount of experimental data under a wide range of operating conditions, and predicting degradation is challenging or impossible.

Alternatively, physics-based models describing the thermodynamics, reaction kinetics and transport within the cell are valid over a wide range of operating conditions and could be coupled to degradation models directly. Physics-based models have been widely used for battery design \cite{Doyle1993, Doyle1995, Fuller1994, Doyle2003} but are usually too computationally intensive for the limited resources of an embedded BMS. The so-called pseudo two-dimensional (P2D) model developed by the Newman group \cite{Doyle1993} is probably the most widely used lithium-ion battery model of this type. It is composed of a one-dimensional macro-scale model describing the evolution of lithium concentration and electric potential in the electrolyte across the anode, separator and cathode and micro-scale models for the electrodes. The pseudo-second dimension arises from these coupled one-dimensional micro-scale models describing the solid-phase diffusion of lithium in the porous active material of the electrodes. These micro-scale models solve the diffusion of lithium occurring in a spherical particle at each local position of the macro-scale porous electrode model. By modelling diffusion and kinetics limitations, the P2D model is able to accurately describe lithium-ion battery dynamics over a wide operating range \cite{Forman2012} and is therefore an excellent starting point for the next generation of BMSs.

However, the computation required by the P2D model is intense compared to ECMs for embedded applications. Several attempts at performing state estimation on simplified models derived from the P2D model have been reported in the literature. A common simplification known as the single particle model (SPM), assumes that each electrode can be represented by a unique solid-phase particle and neglects concentration gradients in the electrolyte. State estimation using the SPM and similar approximations has already been reported in the literature, and includes the use of an extended Kalman filter (EKF) algorithm \cite{Santhanagopalan2006a}, or a backstepping PDE state estimator \cite{Moura2013b}. In \cite{DiDomenico2010}, the EKF was applied to an averaged electrochemical model similar to the SPM to estimate SOC and critical concentration at the surface of the electrodes. However, these approaches are inherently limited due to the low current validity range of the SPM.
Other approaches include state estimation on reduced-order models derived from the P2D model. In \cite{Smith2008b,Smith2010}, Kalman filtering is performed on a reduced-order state variable model computed by residue grouping \cite{Smith2007b,Smith2008} from transcendental transfer functions approximating each equation of the P2D model assuming quasi-linear behaviour. In \cite{Stetzel2015a}, the EKF is applied to a state space reduced-order model computed from the P2D model using a discrete-time realization algorithm \cite{Lee2012a,Lee2012b,Lee2014a}. However, the parameters of such reduced-order models may be difficult to interpret or have no direct physical meaning, which makes accounting for degradation effects difficult.

Recent works have shown that using spectral numerical methods instead of the commonly used finite-difference method to discretise the P2D model results in a highly reduced model order whilst maintaining accuracy and physical significance of parameters. Dao et al. used the Galerkin spectral method on sinusoidal basis functions to discretise the electrolyte diffusion equation \cite{Dao2012a}, while Cai and White applied orthogonal collocation on finite elements to all the equations of the P2D model \cite{Cai2012}. Orthogonal collocation enforced at zeros of Jacobi polynomials was also applied to the full P2D model in \cite{Northrop2011b}  and solved using Maple and DASSL solvers using cosine basis functions and more recently Chebyshev polynomial basis functions for improved convergence at high currents \cite{Suthar2014}. In previous work, we applied Chebyshev orthogonal collocation to the isothermal P2D model and showed that computation time could be reduced by a factor of 10 to 100 compared to finite-difference for the same result accuracy \cite{Bizeray2013}. We have also successfully applied this approach for simulation of supercapacitors \cite{Drummond2015}.

In this work, we applied  the EKF algorithm to the thermal-electrochemical P2D model solved using Chebyshev orthogonal collocation for battery state estimation. State estimation of the full P2D model solved using the approach discussed in \cite{Northrop2011b} has recently been reported using the optimisation-based moving-horizon estimation technique \cite{Suthar2013} and a tethered particle filter algorithm \cite{Gopaluni2013}.
However, our approach using the simpler EKF algorithm is much less computationally intensive while showing good performance. To our knowledge, this is the first attempt at estimating the states of the full P2D model using the EKF algorithm.
Solving a high-fidelity model such as the P2D model coupled to degradation models online a BMS can provide valuable information on the internal states of the battery, enabling new safety limits \cite{Chaturvedi2010} and advanced health-conscious control algorithms \cite{Moura2013} to be used.

\section{Thermal-electrochemical model}
\label{sec:thermalElectrochemicalModel}
The battery model considered here consists of the P2D electrochemical model coupled to a bulk thermal model. The electrochemical model describes the transport of lithium, reaction kinetics and thermodynamics at the electrode level while the bulk thermal model describes the evolution of cell temperature. The electrochemical and thermal models are coupled together through the potential- and concentration-dependent heat generation rate and the temperature-dependent physical and chemical properties of the P2D model.

\subsection{Electrochemical model}
\label{subsec:electrochemicalModel}

A lithium-ion cell consists of two porous electrodes composed of an active material that can store lithium intercalated in the solid material, and a separator that allows the passage of ions but not electrons. The electrodes and the separator are soaked in an electrolyte that allows the transport of ions. During discharge, lithium stored in the anode is de-inserted from the active material and released as ions in the electrolyte. Driven by diffusion (concentration gradient) and migration (potential gradient), lithium ions travel through the separator to the cathode where they are inserted in the lattice of the cathode active material. Simultaneously, electrons travel from the anode to the cathode through the external circuit, powering a load, to ensure electro-neutrality. This process is reversed during battery charging.

\begin{figure}
\centering
\includegraphics[width=0.45\textwidth]{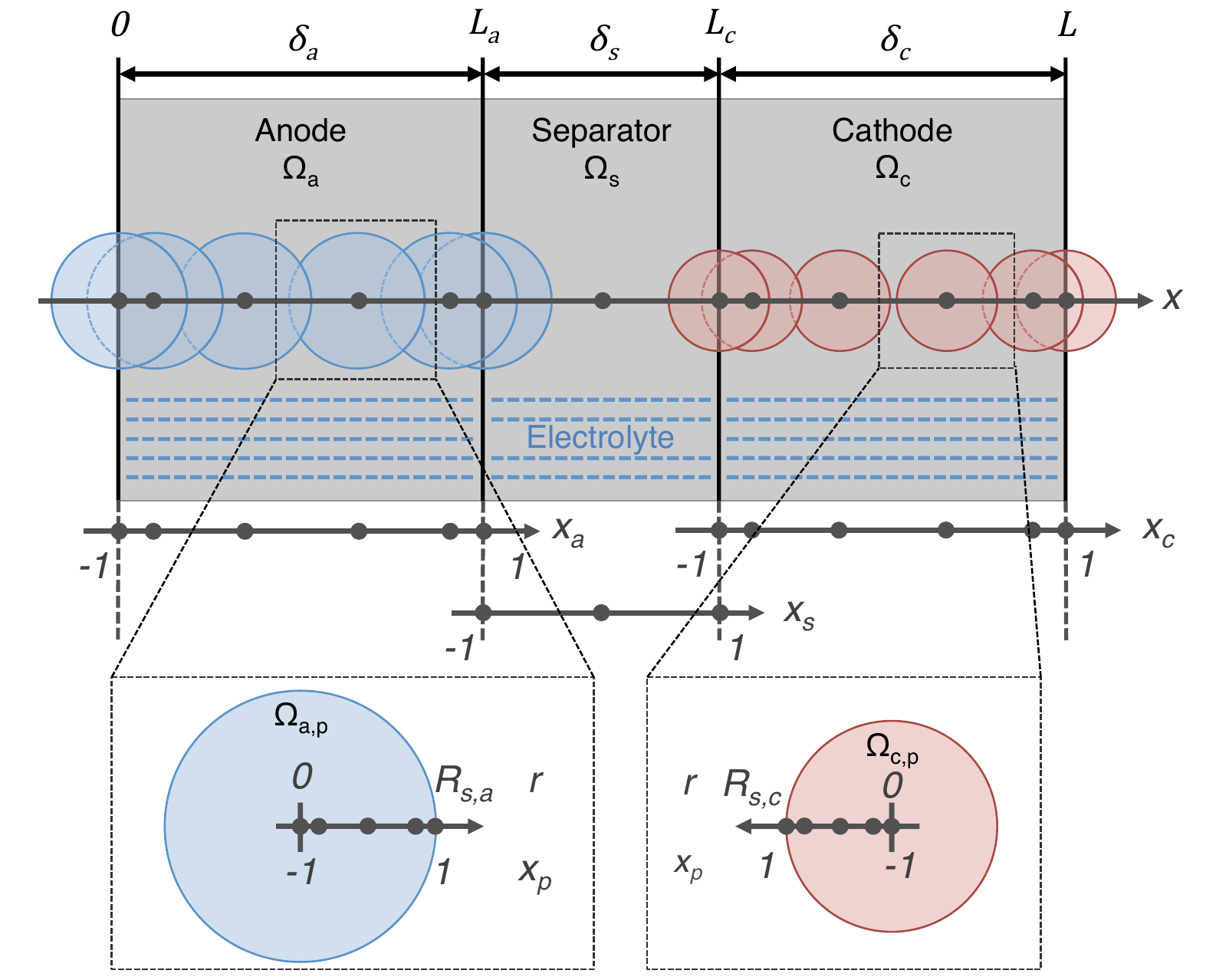}
\caption{Schematic of the cell computational domains. The cell is divided in three domains, anode $\Omega_a$, separator $\Omega_s$ and cathode $\Omega_c$, where two phases are present: the  electrolyte phase and the solid phase. The porous nature of each electrode is considered by assuming spherical particles of solid-phase material $\Omega_{a,p}$ and $\Omega_{c,p}$ at each local position in the anode and cathode domains respectively. The physical coordinate of the domain are the \textit{x}-coordinates across the cell and the radial \textit{r}-coordinates in the particle. The computational domain is rescaled to $\left[ -1, 1 \right]$ and consists of different sets of Chebyshev collocation nodes $x_a$, $x_s$, $x_c$ and $x_p$.}
\label{fig:schematic_cell}
\end{figure}

From a mathematical modelling perspective, the cell is divided into three domains: anode, separator and cathode, denoted $\Omega_a$, $\Omega_s$ and $\Omega_c$ respectively (Fig.~\ref{fig:schematic_cell}). In each of these domains, two phases are considered, the solid phase and the electrolyte phase, and are treated as superimposed continua using porous electrode theory \cite{Newman1975}, sometimes called homogenization, therefore neglecting the exact micro-structure of the electrodes.
In order to account for the tortuosity of the porous material, effective electrolyte diffusivity $D_e^{eff} = D_e \epsilon_e^b$ and ionic conductivity $\kappa^{eff} = \kappa \epsilon_e^b$ are considered with $b$ the so-called Bruggeman coefficient \cite{Doyle1993,Fuller1994,Chung2013}.

The P2D model consists of a set of partial differential equations (PDEs) and algebraic constraints governing the evolution of lithium concentration and electric potential within the cell. 
The dependent variables are solid-phase concentration $c_s(r,x,t)$, electrolyte concentration $c_e(x,t)$, electric potential at the surface of the solid-phase particles $\phi_s(x,t)$, electric potential in the electrolyte $\phi_e(x,t)$ and volumetric reaction current $j^{Li}(x,t)$, which expresses the amount of lithium exchanged between the solid-phase and the electrolyte per unit volume of electrode.
The independent variables are time $t$, the $x$-coordinate across the cell thickness and the spherical $r$-coordinate in the solid-phase particles.
The transport of lithium in each spherical particle is described by the spherical diffusion equation \eqref{eq:P2Dmodel_solidPhase_diffusion} with Neumann boundary conditions \eqref{eq:P2Dmodel_solidPhase_diffusion_bc}:
\begin{gather}
\label{eq:P2Dmodel_solidPhase_diffusion}
\frac{\partial c_{s}}{\partial t} = 
\frac{1}{r^2} \frac{\partial}{\partial r} \left(r^2 D_{s}\frac{\partial c_{s}}{\partial r} \right) \\
\label{eq:P2Dmodel_solidPhase_diffusion_bc}
\left. \frac{\partial c_{s}}{\partial r} \right|_{r=0} =0 \quad 
\text{and} \quad D_{s} \left. \frac{\partial c_{s}}{\partial r}\right|_{r=R_{s}} = \frac{-j^{Li}}{a_{s} \mathcal{F}} 
\end{gather}
The change of variable $\bar{c}_s = r c_s$ is introduced to simplify this sub-model implementation (\ref{app:changeOfVar}).
The reaction current $j^{Li}$ is given as a function of the electrode local overpotential $\eta$ by the Butler-Volmer kinetics equation:
\begin{equation}
\label{eq:P2Dmodel_BVkinetics}
j^{Li} = a_s i_0 \left[ \exp \left( \frac{\alpha_a \mathcal{F}}{RT} \eta \right) - \exp \left( \frac{-\alpha_c \mathcal{F}}{RT} \eta \right) \right]
\end{equation}
where $a_s = 3 \epsilon_s / R_s$ is the specific interfacial area of the electrode. The exchange current density $i_0$ depends on the particle surface concentration $c_s^{surf}$ and the electrolyte concentration $c_e$ according to:
\begin{equation}
\label{eq:P2Dmodel_exchangeCurrent}
i_{0} = k \mathcal{F} \left( c_s^{max} - c_s^{surf} \right)^{\alpha_a} \left( c_s^{surf} \right)^{\alpha_c} \left( c_e \right)^{\alpha_a}
\end{equation}
The overpotential in \eqref{eq:P2Dmodel_BVkinetics} is given by $\eta = \phi_s - \phi_e - U (c_s^{surf})$ where the experimentally-fitted open-circuit potential functions $U(c_s^{surf})$ are taken from \cite{Ramadass2004}.

The evolution of lithium concentration in the electrolyte is governed by the diffusion equation \eqref{eq:P2Dmodel_electrolyte_diffusion} subject to homogeneous Neumann boundary conditions.
\begin{gather}
\label{eq:P2Dmodel_electrolyte_diffusion}
\epsilon_e \frac{\partial c_e}{\partial t} = \frac{\partial }{\partial x} \left( D_e^{eff} \frac{\partial c_e}{\partial x} \right)
+ \frac{1-t^0_+}{\mathcal{F}} j^{Li} \\
\end{gather}
The electrolyte potential $\phi_e$ is governed by Ohm's law:
\begin{equation}
\label{eq:P2Dmodel_electrolyte_potential}
\kappa^{eff} \frac{\partial \phi_e}{\partial x} - 
\kappa^{eff}_D \frac{\partial \ln c_e}{\partial x} + i_e = 0
\end{equation}
where the diffusional conductivity is given by:
\begin{equation}
\label{eq:P2Dmodel_diffusionalElectrolyteConductivity}
\kappa^{eff}_D = \frac{RT}{\mathcal{F}}   \left(1 - 2t^0_+ \right)  \kappa^{eff} 
\end{equation}
It has recently been reported \cite{Ramos2015} that mistakes are sometimes made in the literature regarding \eqref{eq:P2Dmodel_electrolyte_potential} and \eqref{eq:P2Dmodel_diffusionalElectrolyteConductivity} and we therefore are careful to use the correct expressions here. The electrolyte potential at the cathode current collector is chosen as the reference potential and set to zero to ensure that the system of equations is fully constrained.
The solid phase potential at the surface of the particles is governed by Ohm's law:
\begin{equation}
\label{eq:P2Dmodel_solidPhase_potential}
\sigma^{eff} \frac{\partial \phi_s}{\partial x}  + i_s = 0
\end{equation}
The local fractions of current density carried by the ions in electrolyte $i_e$ and electrons in the solid-phase $i_s$ are related to the total current density passing through the cell $i_{app}$ by the Kirchoff's law $i_s + i_e = i_{app}$.
By virtue of conservation of charge, the local reaction rate $j^{Li}$ is equal to the divergence of the electrolyte current density.
The input of the model is the current $I$ from which the applied current density $i_{app} = I/A_s$ is calculated knowing the electrode surface area $A_s$. The terminal voltage of the cell $V$ is equal to the difference between the solid-phase potential at the cathode current collector and that at the anode current-collector, minus the ohmic drop due to the contact resistance $R_c = 20~\Omega.cm^2$ \cite{Smith2006b} at the current collector/electrode interfaces.

The cell considered for this study consists of a lithium cobalt oxide (LiCoO\textsubscript{2}) cathode and a mesocarbon microbead (MCMB) anode with 1M LiPF\textsubscript{6} in propylene carbonate, ethylene carbonate and dimethyl carbonate (PC:EC:DMC) electrolyte.
The parameters were found in the literature and are summarised in Table~\ref{tab:electrochemical_model_parameters}. It has been shown that electrolyte properties are highly dependent on lithium concentration and cell temperature \cite{Valoen2005}. The empirical expressions for diffusivity $D_e$ and ionic conductivity $\kappa$ as a function of concentration and temperature reported in \cite{Valoen2005} were used in the present work and the transference number $t_0^+ = 0.435$ is assumed constant \cite{Kumaresan2008}.
The focus of the present work is on the efficient solution and state estimation of the P2D model. We acknowledge that the estimation of model parameters from experimental data is crucial for the practical implementation of a state estimation algorithm. This is a challenging task due to the large number of parameters compared to the limited number non-invasive measurements available and this will be the focus of future work.

\begin{table*}[ht]
\caption{Set of parameters of the electrochemical P2D model used for the simulations}
\label{tab:electrochemical_model_parameters}
\centering		                    
\begin{tabular}{l c c c c c}        
\hline							    
Parameter & Units & Anode & Separator & Cathode  & Ref.\\
& & Li\textsubscript{x}C\textsubscript{6} & LiPF\textsubscript{6} & Li\textsubscript{y}CoO\textsubscript{2} & 	\\
\hline							    
$\delta_{i}$ & $ \mu m$ & $73.5$ & $25.0$ & $70.0$ & \cite{Cai2012} \\
$R_{i}$ & $ \mu m$ & $12.5$ & $-$ & $8.5$ & \cite{Cai2012} \\
$\epsilon_{i}$ & $-$ & $0.4382$ & $0.45$ & $0.3$ & \cite{Kumaresan2008} \\
$\epsilon_{f,i}$ & $-$ & $0.0566$ & $-$ & $0.15$ & \cite{Kumaresan2008} \\
$\alpha_{i}$ & $-$ & $0.5$ &  $-$ & $0.5$ & \cite{Doyle2003} \\
$k_{i}^{ref}$ & $m^{2.5}.mol^{-0.5}.s^{-1}$ & $1.764 \times10^{-11}$ & $-$ &  $6.667\times10^{-11}$ & \cite{Kumaresan2008} \\
$D_{s,i}$ & $m^{2}.s^{-1}$ & $5.5\times 10^{-14}$& $-$ & $1.0 \times 10^{-11} $ & \cite{Doyle2003}\\
$\sigma_{i}$ & $S.m^{-1}$ & $100$ & $-$ & $10$ & \cite{Doyle2003} \\
$b_{i}$ & $-$ & $4.1$ & $2.3$ & $1.5$ & \cite{Kumaresan2008} \\
$c_{s,i}^{max}$ & $mol.m^{-3}$ & $30555$ & $-$ & $51555$ & \cite{Ramadass2004} \\
$\theta_{i}^{0}$ & $-$ & $0.756$ & $-$ & $0.465$ & \cite{Cai2012} \\
\hline												
\end{tabular}
\end{table*}

\subsection{Lumped thermal model}
\label{subsec:thermalModel}
The P2D electrochemical model is coupled to a lumped thermal model described by the following energy balance equation:
\begin{equation}
\label{eq:heatEquation}
\rho c_p \frac{dT}{dt} = \dot{q}_{gen} + \dot{q}_{conv}
\end{equation}
The total heat generation rate per unit volume $\dot{q}_{gen}$ is assumed uniform and attributed to four main contributions according to $\dot{q}_{gen} = \dot{q}_{rxn} + \dot{q}_{rev} + \dot{q}_{ohm} + \dot{q}_{c}$, where $\dot{q}_{rxn}$ is the reaction heat generation rate, $\dot{q}_{rev}$ is the reversible heat generation rate due to entropy changes in the active material of electrodes during the intercalation/de-intercalation of lithium and $\dot{q}_{ohm}$ is the electronic and ionic ohmic heat generation rate due to the motion of lithium. The average heat generated by each of these per unit volume \cite{Kumaresan2008,Wu2013c} is given by equations \eqref{eq:heatGeneration_rxn}-\eqref{eq:heatGeneration_rev}-\eqref{eq:heatGeneration_ohm}:
\begin{align}
\label{eq:heatGeneration_rxn}
\dot{q}_{rxn} ={}& \frac{1}{L} \int_{0}^{L} j^{Li} \left( \phi_s - \phi_e - U^{ocp} \right) dx \\
\label{eq:heatGeneration_rev}
\dot{q}_{rev} ={}& \frac{1}{L} \int_{0}^{L} j^{Li} \left( T \frac{\partial U^{ocp}}{\partial T}\right)dx \\
\begin{split}
\label{eq:heatGeneration_ohm}
\dot{q}_{ohm} ={}& \frac{1}{L} \int_{0}^{L} \left[ \sigma^{eff} \left( \frac{\partial \phi_s}{\partial x} \right)^2 + \kappa^{eff} \left( \frac{\partial \phi_e}{\partial x} \right)^2 \right. \\ & \qquad \qquad \qquad \left. + \kappa^{eff}_{D} \left(\frac{\partial \ln c_e }{\partial x} \right) \left(\frac{\partial \phi_e}{\partial x}\right) \right] dx
\end{split}
\end{align}
The ohmic heat generated per unit volume $\dot{q}_{c}$ due to the contact resistance between the electrodes and current collectors is given by:
\begin{equation}
\label{eq:heatGeneration_contact}
\dot{q}_c = \frac{R_c}{A_s V_c} I^2
\end{equation}
The rate of convective heat removal per unit volume from the cell to the coolant air $\dot{q}_{conv}$ in \eqref{eq:heatEquation} is given by:
\begin{equation}
\label{eq:heatRemoval_convection}
\dot{q}_{conv} = - \frac{h A_c \left(T - T_{\infty} \right)}{V_c}
\end{equation}
In this work, it has been assumed that the cell is an 18650 cylindrical cell and therefore the ratio $A_c/V_c = 253$~m\textsuperscript{-1}.
Other cell geometries can easily be considered since only the convective surface area to cell volume ratio is required in this model. However, for large cells the assumption of uniform cell temperature may not be satisfactory as large temperature gradients build up within the cell.

During high C-rate operation, the cell temperature can significantly increase and affect the cell physical and chemical properties. Therefore, the coupling between the thermal and electrochemical model must include the temperature dependency of the model parameters.
Temperature dependencies of the electrolyte diffusivity and conductivity are taken from  \cite{Valoen2005}.
The solid phase diffusion coefficient $D_s$ and the reaction kinetics constant $k$ are also highly dependent on temperature. A common approach assumes an Arrhenius' law temperature dependency given by \cite{Guo2011}:
\begin{equation}
\label{eq:arrheniusLaw}
\psi = \psi^{ref} \exp \left[ \frac{E_a^{\psi}}{R} \left( \frac{1}{T^{ref}} - \frac{1}{T} \right) \right]
\end{equation}
where $\psi$ denotes the parameter considered and $\psi^{ref}$ is the value of this parameter at $T^{ref}$.
Temperature also has an impact on the open-circuit potential of the electrodes. In this paper, this was approximated using a first-order Taylor series expansion with respect to temperature:
\begin{equation}
\label{eq:temperatureDependentOCP}
U = U^{ref} + \left(T-T^{ref} \right) \left(\frac{\partial U  }{\partial T} \right)
\end{equation}
where $U^{ref}$ is the open-circuit potential at $T^{ref}$ and $\left( \partial U/\partial T \right)$ is the entropy change coefficient. Empirical expressions reported in \cite{Guo2011,Egorkina1998a} for the entropy change as a function of solid-phase surface stoichiometry of LiCoO\textsubscript{2} and MCMB electrodes were used for the simulations.
The parameters of the thermal model are summarised in Table~ \ref{tab:thermal_model_parameters}.

\begin{table}
\caption{Thermal model parameters}
\label{tab:thermal_model_parameters}
\centering	
\begin{tabular}{l c c c }
\hline
Parameter & Units & Value & Ref. \\
\hline
$c_p$ & $J.kg^{-1}.K^{-1}$ & $750$ & \cite{Guo2011} \\
$\rho$ & $kg.m^{-3}$ & $1626$ & \cite{Guo2011} \\
$T^{ref}$ & $K$ & $298$ & \cite{Guo2011} \\
$h$ & $W.m^{-2}.K^{-1}$ & $30$ & - \\
$T_{\infty}$ & $K$ & $298$ & - \\
$E_a^{D_{s,1}}$ & $kJ.mol^{-1}$ & $35$ & \cite{Guo2011,Kulova2006}\\
$E_a^{D_{s,3}}$ & $kJ.mol^{-1}$ & $29$ & \cite{Guo2011,Nakamura2000}\\
$E_a^{k_{1}}$ & $kJ.mol^{-1}$ & $20$ & \cite{Guo2011,Egorkina1998a} \\
$E_a^{k_{3}}$ & $kJ.mol^{-1}$ & $58$ & \cite{Guo2011,Zheng2005}\\
\hline
\end{tabular}
\end{table}

\subsection{Chebyshev orthogonal collocation}
\label{subsec:orthogonalCollocation}

The thermal-electrochemical P2D model consists of a set coupled nonlinear partial-differential equations in time and space. An analytical solution for such a complex problem is not available and numerical methods are employed to spatially discretise the equations in the \textit{x}- and \textit{r}-directions. The discretised P2D model consists of a system of ODEs and DAEs that can be integrated using a standard time-adaptive ODE/DAE solver such as MATLAB's \emph{ode15s} \cite{Shampine1999}.
The finite difference method has been commonly used to discretise the P2D model in space but this requires a significant number of discrete nodes and therefore results in a large system of equations. In this paper, the electrochemical P2D model is discretised using a class of spectral methods called Chebyshev orthogonal collocation that results in a much smaller system of equations compared to finite difference for a similar accuracy \cite{Trefethen2000}.

Spectral methods consist of expanding the solution $u$ of a differential equation in terms of chosen orthogonal basis functions and determining the coefficients of this expansion to satisfy the differential equation.
For problems with periodic boundary conditions, cosine functions are a natural choice of basis functions. However, for non-periodic boundary conditions, discontinuities introduced at the boundaries result in Gibbs phenomena that drastically impede spectral accuracy. This can be circumvented by adding linear and/or quadratic terms to the Fourier series expansion to enforce the boundary conditions as in \cite{Northrop2011b}.
However, Chebyshev polynomials are a more natural choice for the solution of differential equations with non-periodic boundary conditions such as the P2D model \cite{Gottlieb1977,Trefethen2000}. The solution $u(x,t)$ of the PDE is therefore approximated by the truncated Chebyshev expansion:
\begin{equation}
\label{eq:truncatedChebyshevExpansion}
u_N(x,t) = \sum_{k = 0}^{N} \hat{u}_{k}(t) T_{k}(x), \quad x \in \left[-1,1\right]
\end{equation}
where $\hat{u}_k(t)$ are the $N+1$ Chebyshev coefficients of the expansion that need to be determined and $T_k$ denotes the Chebyshev polynomial of the first-kind of degree $k$.
In the present work, the coefficients are determined by the so-called orthogonal collocation method, also sometimes referred to as the pseudo-spectral method.
The coefficients $\hat{u}_k(t)$ are calculated by forcing the truncated Chebyshev series \eqref{eq:truncatedChebyshevExpansion} to satisfy the differential equation exactly at the discretising nodes $x_i$ given by:
\begin{equation}
\label{eq:collocationNodes}
x_i = \cos \left( \frac{i \pi}{N} \right) \quad i = 0,1,...,N
\end{equation}
By choosing the coefficients $\hat{u}_k(t)$ so that $u_N (x_i,t) = u(x_i,t)$, the Chebyshev series expansion \eqref{eq:truncatedChebyshevExpansion} becomes an interpolating polynomial of degree $N$ to the solution $u$ of the differential equation at the collocation nodes $x_i$.
It can be shown \cite{Weideman2000} that the interpolating polynomial can be expressed in terms of the value of the solution at the collocation nodes $u_j(t) = u(x_j,t) = u_N(x_j,t)$ by:
\begin{equation}
\label{eq:interpolantChebyshevSeries}
u_N(x,t) = \sum_{j=0}^N u_j(t) \phi_j(x) , \quad x \in \left[-1,1\right]
\end{equation}
where the functions $\phi_j$ are given by:
\begin{equation}
\phi_j(x) = \frac{(-1)^{j+1} (1-x^2)T'_N(x)}{\bar{c}_j N^2 (x-x_j)}, \quad x \in \left[-1,1\right]
\end{equation}
with $\bar{c}_j = 2$ for $j = 0$ and  $j = N$ and $\bar{c}_j = 1$ otherwise.
When implementing the orthogonal collocation method for the solution of PDEs, the coefficients $\hat{u}_k$ of the series expansion are rarely computed explicitly but the differentiation of $u$ is usually performed using a differentiation matrix.
The $p$th derivative of the solution $u$ with respect to $x$ evaluated at the collocation points can be expressed by:
\begin{equation}
u^{(p)}_N (x_i) = \sum_{j=0}^N d_{i,j}^{(p)} u_N(x_j), \quad i = 0,1,...,N
\end{equation}
where the coefficients $d_{i,j}^{(p)}$ can be determined by evaluating the $p$th derivative of the interpolant \eqref{eq:interpolantChebyshevSeries} at the collocation nodes \eqref{eq:collocationNodes}. The coefficients $d_{i,j}^{(p)}$ are the elements of the so-called differentiation matrix $D^p$, which is the discrete approximation to the $p$th derivative operator $\partial^p/\partial x^p$. The derivative of $u$ evaluated at the collocation nodes $\mathbf{u}^{(p)}$ can be expressed in terms of the value of $u$ at the collocation nodes $\mathbf{u}$ with:
\begin{equation}
\mathbf{u}^{(p)} = D^p \mathbf{u}
\end{equation}
The MATLAB function \emph{chebdif.m} discussed in \cite{Weideman2000} was used to compute the Chebyshev orthogonal collocation differentiation matrices.

The P2D model must satisfy the boundary conditions associated with the PDEs as discussed in Section~\ref{subsec:electrochemicalModel}.
In the present work, these boundary conditions are accounted for by reducing the size of the differentiation matrix, since each boundary condition gives an additional constraint that can be used to express the value of the solution at a chosen collocation node in terms of the solution values at all the other collocation nodes. This reduces the size of the differentiation matrix by one row and one column for each boundary condition considered. This leads to reduced differentiation matrices that automatically satisfy the boundary conditions.

\subsection{Domain decomposition and scaling}
\label{subsec:domainDecompositionAndScaling}
The main advantage of spectral methods over FDM is their fast rate of convergence \cite{Trefethen2000}, which means that the same accuracy can be obtained with fewer discretisation nodes (reduced by a factor of 10 to 100).
However, this rapid convergence behaviour, referred to as \emph{spectral accuracy}, is achieved provided that the solution is sufficiently smooth.
Discontinuities reduce spectral accuracy and appear in the electrochemical P2D model at each of the electrode/separator interfaces.
In order to avoid these discontinuities, the cell domain was decomposed into three sub-domains denoted $\Omega_a$, $\Omega_s$ and $\Omega_c$ for the anode, separator and cathode sub-domains respectively, where the model equations are solved on distinct sets of Chebyshev collocation nodes $x_a$, $x_s$ and $x_c$ (Fig.~\ref{fig:schematic_cell}). Additional interface boundary conditions are required to ensure the continuity of the dependent variables and the conservation of flux at the interfaces between sub-domains and are summarised in~\ref{app:interfaceBC_domainDecomposition}. 
The anode and cathode solid-phase particles domains $\Omega_{p,a}$ and $\Omega_{p,c}$ respectively are discretised using the same set of Chebyshev collocation nodes $x_p$.
The number of Chebyshev collocation nodes used in the anode, separator, cathode and particles sub-domains are denoted $N_a$, $N_s$, $N_c$, and $N_p$ respectively.
Each of these sub-domains are rescaled to the interval $[-1,1] \in \mathbb{R}$, since the Chebyshev collocation nodes are defined on this interval.

\subsection{State-space representation of the discretised model}
\label{subsec:DAEstateSpaceRepresentation}
The P2D model discretrised by orthogonal collocation consists of a set of non-linear differential-algebraic equations (DAEs) with respect to time. Using a state-space representation, the model can be conveniently written as a semi-explicit DAE system, consisting of a set of differential \eqref{eq:stateSpaceP2D_diffEq} and algebraic equations  \eqref{eq:stateSpaceP2D_algEq}:
\begin{align}
\dot{\mathbf{x}} &= \mathbf{f} \left( \mathbf{x} , \mathbf{z} , u \right) \label{eq:stateSpaceP2D_diffEq} \\
\mathbf{0} &= \mathbf{g} \left( \mathbf{x} , \mathbf{z} , u \right) \label{eq:stateSpaceP2D_algEq}
\end{align}
where the functions $\mathbf{f}$ and $\mathbf{g}$ are non-linear mapping functions derived from the discretised model equations.
The state vector $\mathbf{x} \in \mathbb{R}^{n_x}$ associated with the differential equations contains the value at the collocation points of the solid-phase concentration $\bar{c}_s = r c_s$ and the electrolyte concentration $c_e$, as well as the bulk temperature~$T$:
\begin{equation}
\mathbf{x} = \left[ \bar{\mathbf{c}}_{s},\mathbf{c}_e , T \right]^{T}
\label{eq:differentialStateVector}
\end{equation}
The state vector $\mathbf{z} \in \mathbb{R}^{n_z}$ associated with the algebraic equations contains the value of the volumetric reaction rate $\mathbf{j}^{Li}$ at the collocation points and the solid-phase electric potential at the cathode and anode current collector $ \phi_{s,c}^0$ and $ \phi_{s,a}^0$ respectively.
\begin{equation}
\mathbf{z} = \left[ \mathbf{j}^{Li} , \phi_{s,c}^0 , \phi_{s,a}^0 \right]^{T}
\label{eq:algebraicStateVector}
\end{equation}
The measurement vector $\mathbf{y} = \left[ V \quad T \right]^T$ containing the value of the voltage $V$ and the temperature $T$ is computed from the differential and algebraic state vectors according to the measurement equation \eqref{eq:stateSpaceP2D_measEq}.
The input $u$ is a scalar equal to the applied current $I$.
\begin{equation}
\mathbf{y}
= \left[ H_x \quad H_z \right] \left[  \begin{matrix} \mathbf{x}\\ \mathbf{z} \end{matrix} \right]
+ H_u u
\label{eq:stateSpaceP2D_measEq}
\end{equation}
The derivation of matrices $H_x$, $H_z$ and $H_u$ is trivial since the temperature is a differential state of the model and the voltage computation is straightforward from the algebraic state vector and the input vector.

Equations \eqref{eq:stateSpaceP2D_diffEq}, \eqref{eq:stateSpaceP2D_algEq} and \eqref{eq:stateSpaceP2D_measEq} constitute a state-space representation of the thermal-electrochemical P2D model. This representation is particularly convenient from a control engineering perspective and can be implemented in the ODE/DAEs MATLAB solver \emph{ode15s}. In addition, this representation is useful for the design and implementation of a state estimator as discussed in Section~\ref{sec:stateEstimation}.

\section{Thermal-electrochemical model simulation results and discussion}
\label{sec:modelResultsAndDiscussion}

In this section,  we compare the model prediction obtained from the solution of the P2D model solved using the Chebyshev orthogonal collocation method discussed in Section~\ref{subsec:orthogonalCollocation} to the solution obtained using the commercial finite-element software COMSOL Multiphysics.
The implementation of the thermal-electrochemical P2D model in COMSOL was performed using the equation-based modelling toolbox similar to \cite{Cai2011a}, see \ref{app:comsolImplementation}.
The P2D model solved by finite-elements in COMSOL is subsequently referred to as the \lq high-fidelity\rq model for simplicity.

\begin{figure}
\centering
\includegraphics[width=0.45\textwidth]{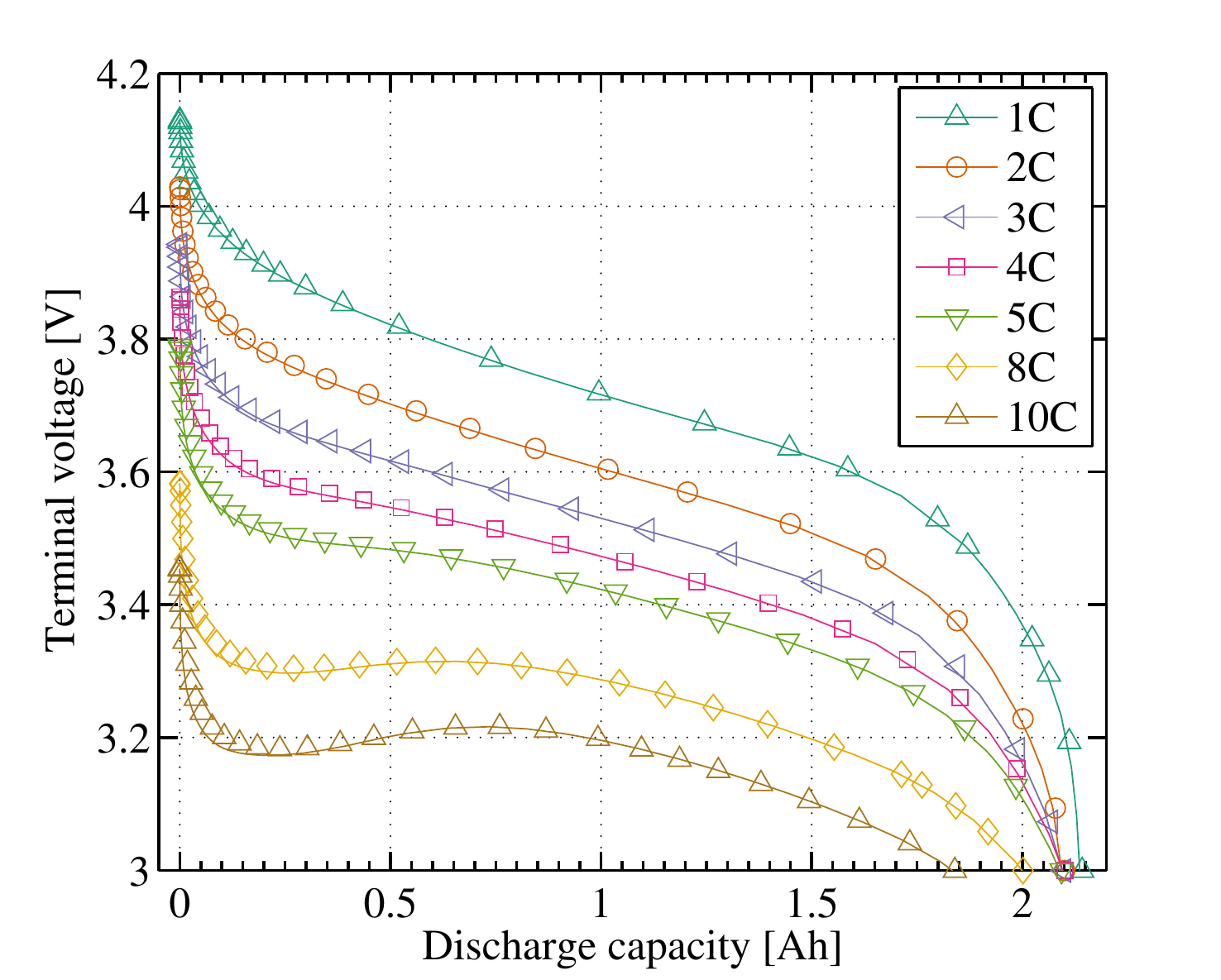}
\caption{Cell voltage under constant-current discharge at several C-rates. Solid lines: COMSOL, markers: Chebyshev orthogonal collocation in MATLAB.}
\label{fig:Voltage_dischargeCurves}
\end{figure}

Fig.\ \ref{fig:Voltage_dischargeCurves} compares the cell terminal voltage predicted by both approaches during constant current discharge at several C-rates ranging from 1C to 10C.
The chosen number of collocation nodes in the anode, separator and cathode are $N_a = 6$, $N_s = 3$ and $N_c = 6$ respectively and the number of collocation nodes in each particles of both electrodes is $N_p = 15$.
The voltage predicted by our approach is in very good agreement with the high-fidelity model up to high C-rates (10C) with a root-mean square and a maximum error of 10~mV and 50~mV respectively.
The solution of the model using Chebyshev orthogonal collocation is typically 30 times faster than the solution using COMSOL.
The computation of a single discharge curve on a desktop computer using a 3.40~GHz processor with 8~GB~RAM is performed in about 5~min with COMSOL, compared to 1~s to 10~s with our implementation.
The order of magnitude of these computation times are consistent with results reported in \cite{Northrop2011b} with a Maple solver.

\begin{figure}
\centering
\includegraphics[width=0.45\textwidth]{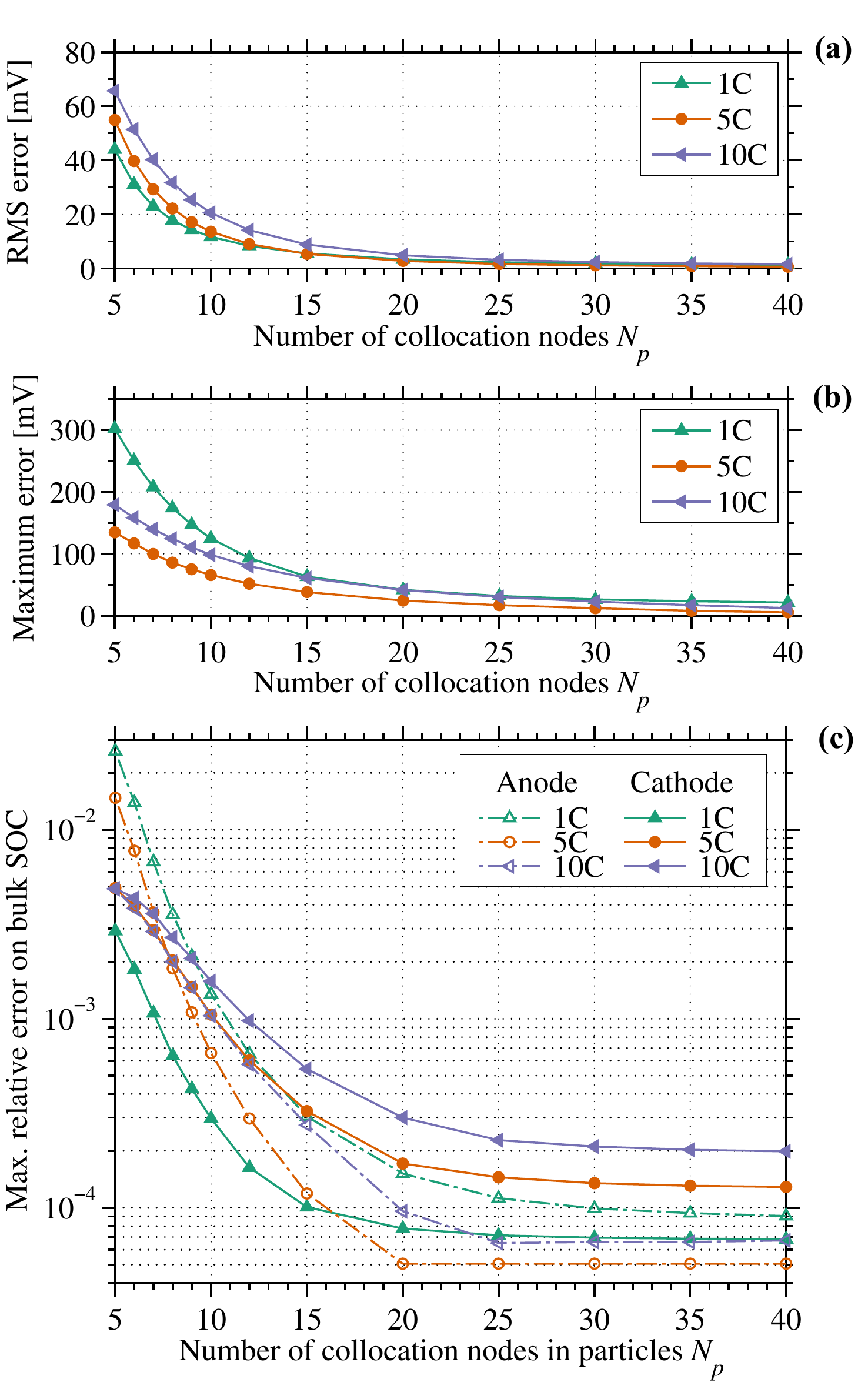}
\caption{RMS error (a) and maximum absolute error (b) on voltage, and maximum relative error on bulk SOC in both electrodes (c) predicted by the P2D model solved using orthogonal collocation in MATLAB compared to the high-fidelity COMSOL model under constant-current discharge at several C-rates with respect to the number of collocation nodes in each solid-phase particles $N_p$.}
\label{fig:Figure3}
\end{figure}

The number of collocation nodes required to discretise the cell domain depends on the C-rate, since higher C-rates result in larger gradients of dependent variables across the cell. In particular, the accuracy of results highly depends on the number of collocation nodes $N_p$ in the solid-phase particles. This is due to the very sharp gradients of lithium concentration at the surface of these particles for medium to high C-rates. The root-mean square and maximum absolute errors between the orthogonal collocation and the high-fidelity model for the 1C, 2C and 5C full constant-current discharge with increasing number of nodes in the solid-phase particles are shown in Fig.~\ref{fig:Figure3}a and Fig.~\ref{fig:Figure3}b respectively. These graphs confirm that a larger number of collocation nodes results in smaller error on voltage prediction, and the higher the C-rate the more nodes are required.
Although, it is suggested by the maximum error graph (Fig.~\ref{fig:Figure3}b) that more collocation nodes are required for the 1C discharge cycle compared to higher C-rate, this is not representative of the whole discharge curve.
The voltage maximum error arises from the very low SOC portion of the discharge curve, when the cell voltage rapidly drops due to the low concentration in the anode material.
The maximum error tends to be smaller at higher C-rates compared to 1C because such a low anode concentration cannot be reached at higher C-rates.

An important state of the model for a BMS is the cell SOC. In this paper, the bulk SOC of the electrode $i$ is defined according to:
\begin{equation}
SOC_i \left(t \right) = \frac{\theta_i^{avg}(t)-\theta_i^{0\%}}{\theta_i^{100\%}-\theta_i^{0\%}}
\label{eq:bulkSOC}
\end{equation}
where $\theta_i^{100\%}$ and $\theta_i^{0\%}$ denote the electrode stoichiometry at 100~\% and 0~\% respectively.
The average electrode stoichiometry $\theta_i^{avg}$ is calculated by integrating the solid-phase concentration in each particle and across the cell according to \eqref{eq:electrodeAverageStoichiometry}.
As shown in Fig.~\ref{fig:Figure3}c, the maximum relative error on the bulk SOC in both the anode and the cathode rapidly falls below 1~\% error with less than 10~nodes in the solid-phase particles and below 0.1~\% with only 15~nodes up to 10C.
\begin{equation}
\theta_s^{avg} \left(t\right)= \frac{3}{ \delta_i R_{s,i}^3} \int_0^{\delta_i} \int_0^{R_{s,i}} r^2 \frac{c_{s,i}\left( x,r,t \right)}{c_{s,i}^{max}} dr dx
\label{eq:electrodeAverageStoichiometry}
\end{equation}

\begin{figure}
\centering
\includegraphics[width=0.45\textwidth]{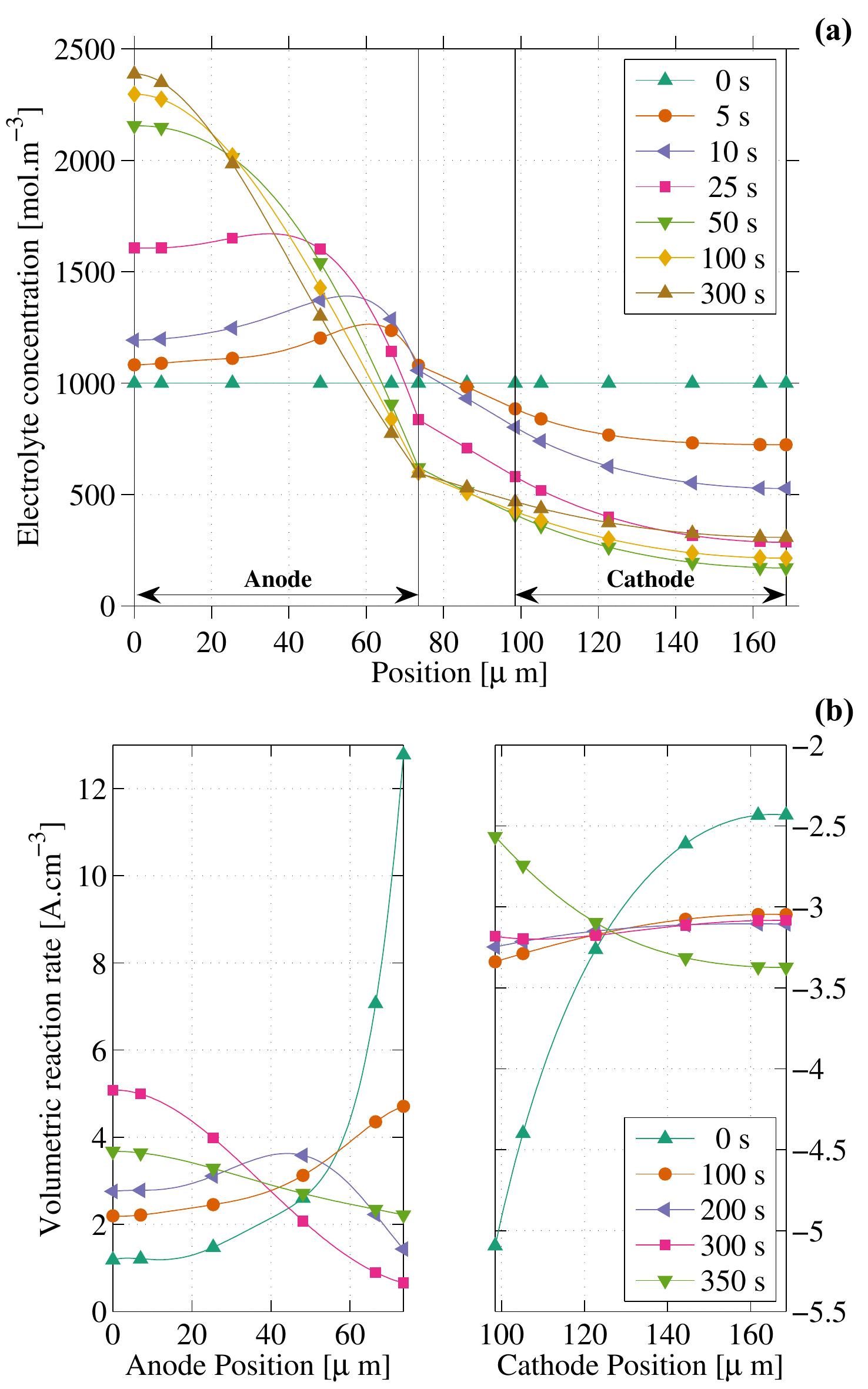}
\caption{Electrolyte concentration (a) and local volumetric reaction rate (b) profiles computed by the thermal-P2D model solved with orthogonal collocation at several time steps estimated under a 10C constant-current discharge.}
\label{fig:Figure4}
\end{figure}

These results confirm the accuracy of our approach. 
Unlike simpler models that have previously been used for battery state estimation, such as the single-particle model, the P2D model is able to predict local variations of internal states across the cell.
Such variations become particularly acute for high C-rate operation (Fig.~\ref{fig:Figure4}), such as discharge where a large amount of lithium is released into the electrolyte at the anode and absorbed at the cathode into the active material.
Due to the relatively slow diffusion of lithium ions from the anode to the cathode, a large concentration gradient builds up in the electrolyte across the cell and reduces the cell performance.
It can be seen from Fig.~\ref{fig:Figure4}a that during a 10C discharge, the electrolyte is almost depleted of lithium in less than 50~s.
The single-particle model also assumes that the insertion reaction rate is uniform within each electrode. Fig.~\ref{fig:Figure4}b shows that this assumption is not valid at high C-rates. At short timescales, the reaction rate at the electrode-separator interface can be an order of magnitude higher compared to the reaction rate at the current collector.
The use of the P2D model for battery state estimation could provide valuable information on local internal states to the BMS and enable the implementation of better health-conscious battery management algorithms \cite{Moura2013}.

\begin{figure}
\centering
\includegraphics[width=0.45\textwidth]{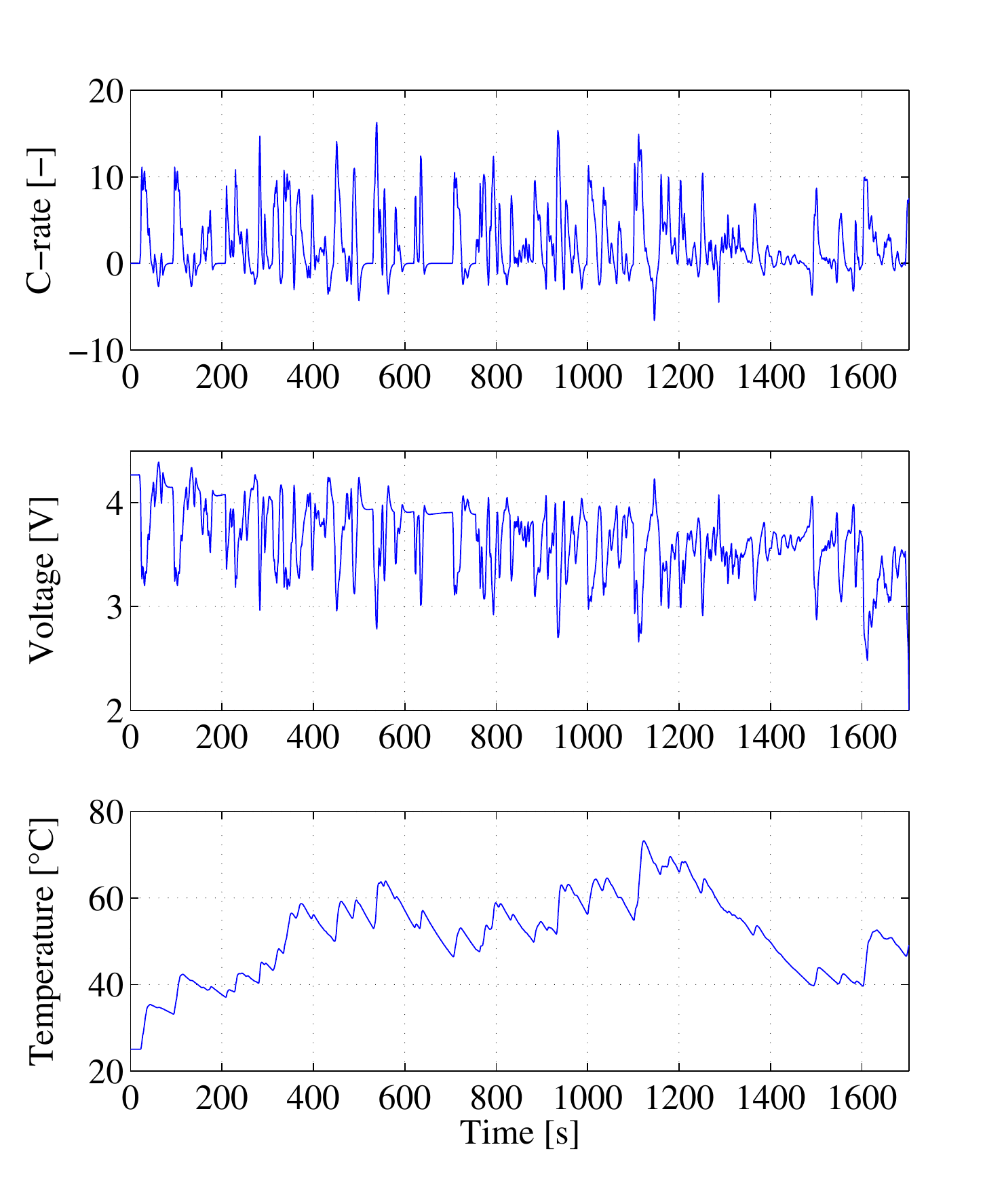}
\caption{Input current (C-rate), voltage and temperature predicted by the thermal-P2D model under a Combined ARTEMIS Driving Cycle.}
\label{fig:CADC_Crate_voltage_Temperature}
\end{figure}

\begin{figure}
\centering
\includegraphics[width=0.45\textwidth]{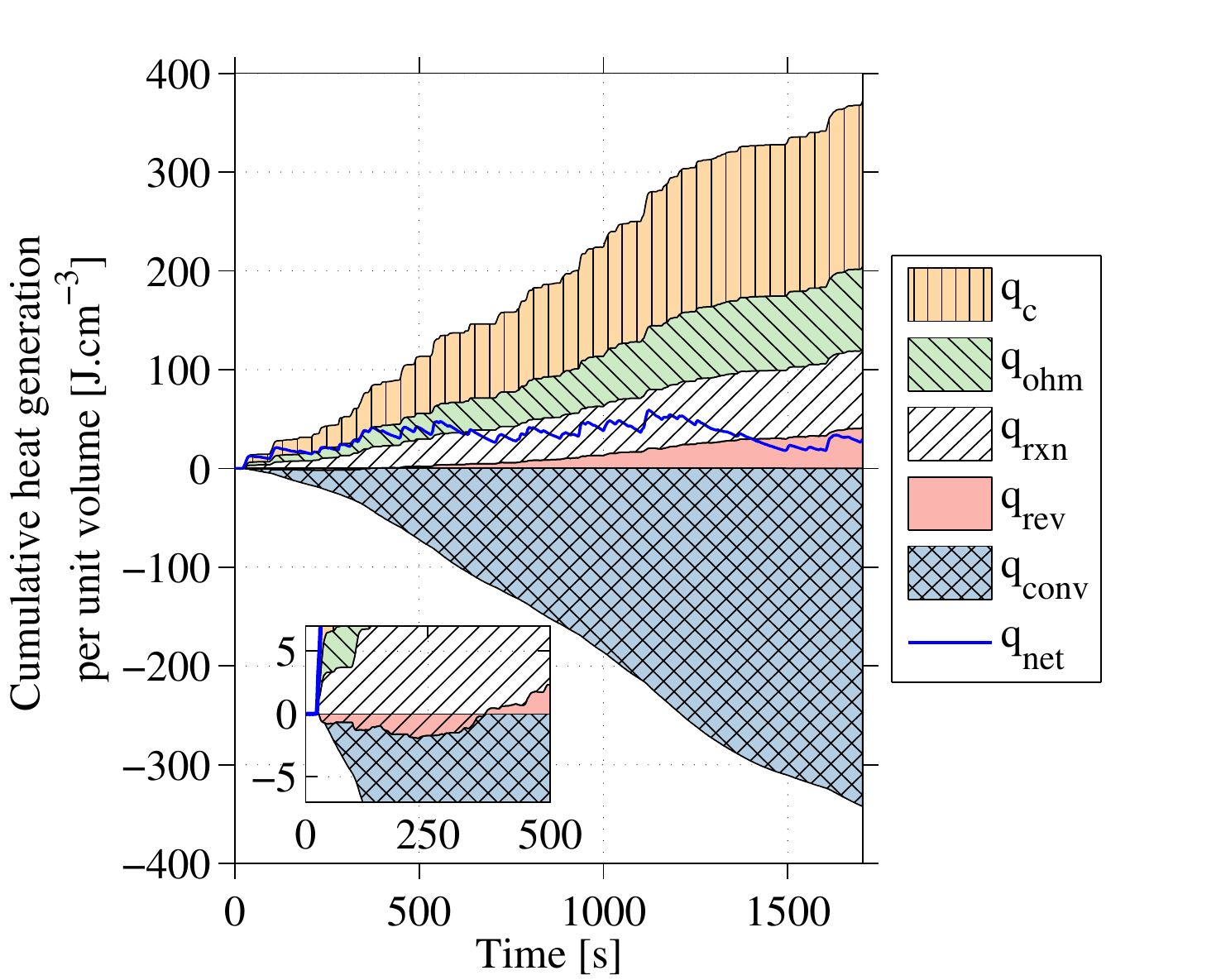}
\caption{Cumulative heat generated and convective heat removed per cell unit volume under the CADC.}
\label{fig:CADC_heat_generation_contributions}
\end{figure}

In embedded application for automotive BMSs, the state estimation would have to be performed with the dynamic current input experienced by the battery pack.
In this work, we used the Combined ARTEMIS Driving Cycle (CADC) \cite{Andre2004} to generate a dynamic current excitation profile approximately representative of an electric car (or PHEV with all-electric mode) drive cycle.
We assumed that the cell input current was proportional to the vehicle's acceleration and that 25~\% of the braking acceleration was recovered to charge the battery.
We chose the scaling factor between car acceleration and input current in order to obtain a relatively aggressive load profile with peak current reaching 15C.
The model prediction of voltage and temperature for the CADC input current are shown in Fig.~\ref{fig:CADC_Crate_voltage_Temperature}.
The model predicts that the full discharge of the battery occurs in 1700~s and that the temperature would rise up to 72~\textdegree C under these relatively aggressive conditions.
The temperature elevation predicted by the model is relatively high compared to what would be experienced by cells in an automotive battery pack because of the high peaks of current and the simplistic air cooling system considered in this study. However, this simulation demonstrates that the model can be solved under highly dynamic and high C-rate operation. The thermal boundary condition could easily be changed to replicate a liquid cooling system if required.
As illustrated by Fig.~\ref{fig:CADC_heat_generation_contributions}, the main contribution to the global heat generation rate arises from the contact resistance heat generation $q_c$ followed by the ohmic heat generation $q_{ohm}$, the reaction heat generation $q_{rxn}$ and the reversible heat generation $q_{rev}$.
Reversible heat is often neglected in the literature for high current operation due to its relatively low magnitude in comparison to other heat sources. However, it can be observed that the reversible heat cumulated over the driving cycle is not negligible.

The numerical solution of the model for a dynamic input with high amplitude current peaks is more intensive than a constant current discharge, and the computation time required by the solver is higher compared to constant-current simulations.
The solution of 1700~s of simulation under the CADC considered required 285~s of computation on the desktop computer previously mentioned.
However, this is still a relatively low computation time since only 168~ms were required on average to solve 1~s of simulation.
This is a promising result for future work on the real-time solution of the thermal-electrochemical model for battery state estimation.

\section{State estimation using a modified EKF}
\label{sec:stateEstimation}

We now discuss the implementation of a Kalman filter for the estimation of the model states \eqref{eq:differentialStateVector} and \eqref{eq:algebraicStateVector} from noisy measurements of $V$, $I$ and $T$.
The Kalman filter is a recursive algorithm that infers the battery internal states by correcting the model states in order to minimise the error between the predicted voltage and temperature and the actual measurements of voltage and temperature for a given current input.

We first give a very brief overview of Kalman filtering for linear models, followed by a detailed explanation of the modifications required for its application to the thermal-electrochemical P2D model. These modifications are motivated by the fact that the P2D model (i) is non-linear and (ii) contains algebraic constraints that cannot be handled by standard algorithms.
In this paper, we applied the extended Kalman filter algorithm for DAEs discussed in~\cite{Becerra2001}  to the thermal-electrochemical P2D model.
The derivation of the Kalman filter equations is not provided in this paper and we refer the reader instead to the literature on Kalman filtering such as~\cite{Gelb1974}.

\subsection{The Kalman filter}
\label{subsec:KalmanFilter}
The Kalman filter is a computationally efficient recursive algorithm for state estimation of dynamic systems described by a stochastic linear state-space models:
\begin{align}
\dot{\mathbf{x}}(t) &= A \mathbf{x}(t) + B u(t) + \mathbf{w}(t)  \label{eq:linearStochasticStateSpaceModel_continuousStateEq} \\
\mathbf{y}_k &= C \mathbf{x}_k + \mathbf{v}_k 
\label{eq:linearStochasticStateSpaceModel_discreteMeasEq}
\end{align}
where \eqref{eq:linearStochasticStateSpaceModel_continuousStateEq} is the continuous state equation governing the system dynamics, and \eqref{eq:linearStochasticStateSpaceModel_discreteMeasEq} is the discrete measurement equation relating the states of the system $\mathbf{x}(t)$ to the available measurements $\mathbf{y}_k$.
The model input $u(t)$ is usually assumed deterministic, while the states and measurements are affected by additive uncorrelated zero-mean Gaussian process noise $\mathbf{w}(t)$ and measurement noises $\mathbf{v}_k$ with covariance matrices $Q$ and $R$ respectively, to account for random environment disturbances and sensor noise.

At every time step $t_k = k T_{s}$, with sampling period $T_s$, an estimate of the state vector $\hat{\mathbf{x}}_k$ and the associated error covariance matrix $\hat{P}_k$ are computed in two steps: the \emph{time update} and the \emph{measurement update}.
In the time update, \emph{a priori} estimates of the state $\hat{\mathbf{x}}^{-}_{k+1}$ and error covariance $\hat{P}^{-}_{k+1}$ at $t_{k+1}$ are calculated using the model and the known input $u(t)$ according to \eqref{eq:linearKalmanFilter_timeUpdate_state} and \eqref{eq:linearKalmanFilter_timeUpdate_errorCov}, where $\Phi = \exp \left(A.T_s\right)$ is the state-transition matrix.
\begin{align}
\hat{\mathbf{x}}_{k+1}^{-} &= \Phi \hat{\mathbf{x}}_{k} + \int_{t_k}^{t_{k+1}} \Phi B \mathbf{u}\left(\tau\right)d\tau \label{eq:linearKalmanFilter_timeUpdate_state} \\
\hat{P}_{k+1}^{-} &= \Phi \hat{P}_{k} + \hat{P}_{k} \Phi^{T} + Q\label{eq:linearKalmanFilter_timeUpdate_errorCov}
\end{align}
In the measurement update, \emph{a posteriori} estimates of the state $\hat{\mathbf{x}}_{k+1}$ and error covariance $\hat{P}_{k+1}$ are computed based on the error between estimated measurements $\hat{\mathbf{y}}^{-}_{k+1}$ and actual noisy measurements $\mathbf{y}_{k+1}$ according to \eqref{eq:linearKalmanFilter_measUpdate_KalmanGain}, \eqref{eq:linearKalmanFilter_measUpdtate_stateCorrection} and \eqref{eq:linearKalmanFilter_measUpdate_errorCovUpdate}.
\begin{align}
K_{k+1} &= \hat{P}_{k+1}^{-} C^{T} \left( C \hat{P}_{k+1}^{-} C^{T} + R \right)^{-1} \label{eq:linearKalmanFilter_measUpdate_KalmanGain} \\
\hat{\mathbf{x}}_{k+1} &= \hat{\mathbf{x}}_{k+1}^{-} + K_{k+1} \left( \mathbf{y}_{k+1} - \hat{\mathbf{y}}_{k+1}^{-} \right) \label{eq:linearKalmanFilter_measUpdtate_stateCorrection} \\
\hat{P}_{k+1} &= \left( I - K_{k+1} C \right) \hat{P}_{k+1}^{-} \label{eq:linearKalmanFilter_measUpdate_errorCovUpdate}
\end{align}
The Kalman filter is the optimal state estimator in the least-squares sense for minimising the state estimation error for linear systems. However, the battery model discussed in Section~\ref{sec:thermalElectrochemicalModel} is non-linear and has algebraic constraints. We therefore discuss a modified version of the Kalman filter based on the EKF algorithm for the state estimation of the battery model.
\subsection{Battery stochastic state-space model}
The EKF algorithm relies on a non-linear stochastic state-space model. Such a state-space representation of the P2D model can be derived from the state-space representation given by \eqref{eq:stateSpaceP2D_diffEq}, \eqref{eq:stateSpaceP2D_algEq} and \eqref{eq:stateSpaceP2D_measEq} by adding process noise and measurement noise to the dynamics and measurement equations respectively according to:
\begin{align}
\dot{\mathbf{x}}(t) &= \mathbf{f} \left(\mathbf{x}(t),\mathbf{z}(t),u(t) \right) + \mathbf{w}(t) \label{eq:nonLinModel_diffEq} \\
\mathbf{0} &= \mathbf{g} \left( \mathbf{x}(t),\mathbf{z}(t) , u(t)\right)
\label{eq:nonLinModel_algEq} \\
\mathbf{y}_k &= \left[ H_x \quad H_z \right]  \left[  \begin{matrix} \mathbf{x}_k\\ \mathbf{z}_k \end{matrix} \right] + H_u  u_k + \mathbf{v}_k \label{eq:nonLinModel_measEq}
\end{align}
Similarly to the Kalman filter discussed in Section~\ref{subsec:KalmanFilter}, the process noise $\mathbf{w}$ and measurement noise $\mathbf{v}_k$ are zero-mean Gaussian additive noises uncorrelated in time with covariance matrices $Q$ and $R$ respectively.
We assumed no noise on the input current $u(t)$.
\subsection{DAE State-space model linearisation}
The difference between the Kalman filter and the EKF consists of additional linearisation steps required between the time update and the measurement update compared to the Kalman filter algorithm for linear models.
The linearisation of the differential equation \eqref{eq:nonLinModel_diffEq} and the algebraic equation \eqref{eq:nonLinModel_algEq} are performed about the current state estimate $[ \hat{\mathbf{x}} , \hat{\mathbf{z}} ]^T$ at every time step \cite{Becerra2001}.
This allows the system of non-linear DAEs to be transformed into a system of locally linear ODEs that can be used in both the time update and measurement update steps discussed in Section~\ref{subsec:KalmanFilter}.

To perform the model linearisation, we first define the following variables:
\begin{align}
\tilde{\mathbf{x}} &= \mathbf{x} - \hat{\mathbf{x}} \\
\dot{\tilde{\mathbf{x}}} &= \dot{\mathbf{x}} - \dot{\hat{\mathbf{x}}} \\
\tilde{\mathbf{z}} &= \mathbf{z} - \hat{\mathbf{z}} \\
\tilde{u} &= u - \hat{u} \\
\tilde{\mathbf{y}} &= \mathbf{y} - \hat{\mathbf{y}}
\end{align}
Assuming that the functions $\mathbf{f}$ and $\mathbf{g}$ are sufficiently differentiable, first-order Taylor series expansions of these functions about the current state estimate are:

\begin{align}
\mathbf{f} \left( \mathbf{x} , \mathbf{z} \right) &= \mathbf{f} \left( \hat{\mathbf{x}} , \hat{\mathbf{z}} \right) + \mathbf{f}_x \tilde{\mathbf{x}} + \mathbf{f}_z \tilde{\mathbf{z}} + \mathbf{f}_u \tilde{u} 
\label{eq:TaylorSeries_diffEq}\\
\mathbf{g} \left( \mathbf{x} , \mathbf{z} \right) &= \mathbf{g} \left( \hat{\mathbf{x}} , \hat{\mathbf{z}} \right) + \mathbf{g}_x \tilde{\mathbf{x}} + \mathbf{g}_z \tilde{\mathbf{z}} + \mathbf{g}_u \tilde{u}
\label{eq:TaylorSeries_algEq}
\end{align}
where $\mathbf{f}_i$ and $\mathbf{g}_i$ denote the partial derivative of $\mathbf{f}$ and $\mathbf{g}$ respectively with respect to the variable $i = \{ \mathbf{x}, \mathbf{z}, u \}$ evaluated at the current state estimate $[ \hat{\mathbf{x}} , \hat{\mathbf{z}} ]^T$. The matrices $\mathbf{f}_x$, $\mathbf{f}_z$, $\mathbf{g}_x$ and $\mathbf{g}_z$ are therefore jacobian matrices of the functions $\mathbf{f}$ and $\mathbf{g}$.
A linear approximation of the DAE system is obtained by substituting \eqref{eq:TaylorSeries_diffEq} and \eqref{eq:TaylorSeries_algEq} into \eqref{eq:nonLinModel_diffEq} and \eqref{eq:nonLinModel_algEq}:

\begin{align}
\dot{\tilde{\mathbf{x}}} &=  \mathbf{f}_x \tilde{\mathbf{x}} + \mathbf{f}_z \tilde{\mathbf{z}} + \mathbf{f}_u \tilde{u} \\
\label{eq:linearized_diffEq}\\
\mathbf{0} &=  \mathbf{g}_x \tilde{\mathbf{x}} + \mathbf{g}_z \tilde{\mathbf{z}} + \mathbf{g}_u \tilde{u}
\label{eq:linearized_algEq}
\end{align}
The random variables $\mathbf{w}$ and $\mathbf{v}_k$ are additive and can therefore be ignored in the linearisation process.

By assuming that the Jacobian matrix $\mathbf{g}_{z}$ is non-singular, which is equivalent to assuming that the semi-explicit DAEs system is of index~1, the linearized algebraic constraint \eqref{eq:TaylorSeries_algEq} can be rearranged to obtain an expression of the algebraic state vector in terms of the differential state vector:
\begin{equation}
\label{eq:linearApproximationAlgebraicState}
\tilde{\mathbf{z}} = - \mathbf{g}_{z}^{-1} \left[  \mathbf{g}_{x} \tilde{\mathbf{x}} + \mathbf{g}_u \tilde{u} \right]
\end{equation}
Substituting \eqref{eq:linearApproximationAlgebraicState} into \eqref{eq:linearized_diffEq} gives the following linearised state equation that includes the algebraic constraint,
\begin{equation}
\dot{\tilde{\mathbf{x}}} = A^{lin} \tilde{\mathbf{x}} + B^{lin} \tilde{u} 
\end{equation}
where:
\begin{align}
A^{lin} &= \mathbf{f}_x - \mathbf{f}_z \mathbf{g}_z^{-1} \mathbf{g}_x
\label{eq:matrix_A_lin}\\
B^{lin} &= \mathbf{f}_u - \mathbf{f}_z \mathbf{g}_z^{-1} \mathbf{g}_u
\label{eq:matrix_B_lin}
\end{align}
In the linearisation process, the DAE system is therefore transformed into an ODE system that can be used in a standard Kalman filter algorithm.
The state-transition matrix $\Phi$ of the linearised model is given by $\Phi = \exp \left( A^{lin} T_s\right)$ and is used in the time update of the Kalman filter discussed in Section~\ref{subsec:KalmanFilter}.
In a similar way, substituting \eqref{eq:linearApproximationAlgebraicState} into the measurement equation~\eqref{eq:nonLinModel_measEq} gives,
\begin{equation}
\label{eq:measurementLinearApproximation}
\tilde{\mathbf{y}} = C^{lin} \tilde{\mathbf{x}} + D^{lin} \tilde{u}
\end{equation}
where:
\begin{align}
C^{lin} &= H_x - H_z \mathbf{g}_z^{-1} \mathbf{g}_x
\label{eq:matrix_C_lin}\\
D^{lin} &= H_u - H \mathbf{g}_z^{-1} \mathbf{g}_u
\label{eq:matrix_D_lin}
\end{align}
The measurement matrix $C^{lin}$ therefore includes the linearised algebraic constraint and can be used in the measurement update of the standard Kalman filter algorithm to compute the Kalman gain and update the error covariance estimate.

\subsection{Summary of modified EKF for non-linear DAEs}
\label{subsec:modifiedEkfForDae}
This section provides a step-by-step description of the modified EKF for systems of DAEs. In \cite{Becerra2001}, the modified EKF algorithm is based on the square-root form of the EKF for numerical stability. This guarantees that the error covariance matrix remains positive semi-definite by using the square-root of the error covariance matrix instead of the error covariance matrix itself.
Although the square-root implementation is more robust in some cases, the two algorithms are mathematically equivalent and we found no difference in the results between the standard and square-root form for our problem. Only the standard version of the modified EKF is discussed in this paper for simplicity. We therefore refer the reader to \cite{Becerra2001} regarding the implementation of the square-root EKF.

The EKF algorithm is initialised by assuming an initial differential state estimate $\mathbf{x}_0$ and error covariance matrix $P_0$ at time $t_0$. A consistent initial algebraic state vector $\mathbf{z}_0$ is computed using the MATLAB solver  \emph{fsolve} for systems of non-linear equations. The computation of an initial error covariance matrix $P_0$ that accounts for the spatial correlation of the model states is crucial for the performance and convergence of the EKF for DAEs. The structure of the matrix $P_0$ must satisfy the spatial correlation of the error, otherwise the conservation of lithium cannot be guaranteed in the measurement update step. For instance, the use of a diagonal initial error covariance matrix (i.e. no spatial correlation) results in the DAE solver failure after a few time steps due to violation of conservation laws. In this work, the structure of the error covariance matrix was obtained numerically using the model states computed by integrating the model under various inputs and initial conditions.

Once the modified EKF for DAEs is initialised, the following algorithm steps are performed recursively:
\begin{enumerate}[1.]

\item
\emph{State time update:}
The current state estimate $[\hat{\mathbf{x}}_k , \hat{\mathbf{z}}_k]$ is projected forward in time to the next time step by integrating the non-linear DAEs system using the MATLAB solver for DAEs \emph{ode15s} from $t_k$ to $t_{k+1}$. The predicted state vector at time $t_{k+1}$ is the \emph{a priori} state estimate $[\hat{\mathbf{x}}_{k+1}^{-} , \hat{\mathbf{z}}_{k+1}^{-}]^T$ at time $t_{k+1}$.

\item 
\emph{Model linearisation:}
The DAE model is linearised about the current state estimate $[\hat{\mathbf{x}}_k , \hat{\mathbf{z}}_k]$ to compute the state transition matrix $\Phi = \exp \left( A^{lin}.T_s \right)$ of the linearised model.

\item
\emph{Error covariance time update:}
The error covariance  $\hat{P}_k$ is propagated in time using \eqref{eq:linearKalmanFilter_timeUpdate_errorCov} to obtain the \emph{a priori} error covariance estimate $\hat{P}_{k+1}^{-}$ at time $t_{k+1}$.

\item
\emph{Model linearisation:}
The DAE model is linearised about the \emph{a priori} state estimate $[\hat{\mathbf{x}}_{k+1}^{-} , \hat{\mathbf{z}}_{k+1}^{-}]^T$ in order to compute the measurement matrix $C^{lin}$ for the measurement update.

\item
\emph{Measurement update:}
The matrix $C^{lin}$ previously computed is used to calculate the Kalman gain according to \eqref{eq:linearKalmanFilter_measUpdate_KalmanGain}.
The \emph{a priori} differential state estimate $\hat{\mathbf{x}}_{k+1}^{-}$ and error covariance estimate $\hat{P}_{k+1}^{-}$ are updated to account for the measurement $\mathbf{y}_{k+1}$ according to \eqref{eq:linearKalmanFilter_measUpdtate_stateCorrection} and \eqref{eq:linearKalmanFilter_measUpdate_errorCovUpdate} respectively. 
The measurement estimate $\hat{\mathbf{y}}_{k+1}$ in \eqref{eq:linearKalmanFilter_measUpdtate_stateCorrection} is computed from the prior estimate $[\hat{\mathbf{x}}_{k+1}^{-} , \hat{\mathbf{z}}_{k+1}^{-}]^T$ and the input $u_{k+1}$ according to the measurement equation \eqref{eq:stateSpaceP2D_measEq}. The \emph{a posteriori} differential state estimate $\hat{\mathbf{x}}_{k+1}$ and error covariance estimate $\hat{P}_{k+1}$ are therefore obtained.
\item
\emph{Consistent algebraic states}
The consistent \emph{a posteriori} algebraic state estimate $\hat{\mathbf{z}}_{k+1}$ is obtained from the posterior differential state estimate $\hat{\mathbf{x}}_{k+1}$ and the input $u_{k+1}$ using the MATLAB \emph{fsolve} function.
\end{enumerate}
This algorithm is repeated recursively at every time step.
\section{State estimation results and discussion}
\label{sec:stateEstimationResults}
In this section, we present simulation results showing the performance of the modified EKF algorithm discussed in Section~\ref{subsec:modifiedEkfForDae} for the state estimation of the full thermal-electrochemical model discussed in Section~\ref{sec:thermalElectrochemicalModel}.
The state estimation of a battery cell requires experimental data as inputs to the EKF, namely the applied current and the measured voltage and temperature response of the cell.
Although the state estimator performance should ultimately be tested against real experimental data, due to the difficulty in verifying \emph{in situ} the internal states in a real battery, we used the thermal-electrochemical model itself to emulate experimental results.
Employing such numerical experiments is worthwhile since the state estimate error can be easily computed.

\begin{figure}[t]
\centering
\includegraphics[width=0.95\textwidth]{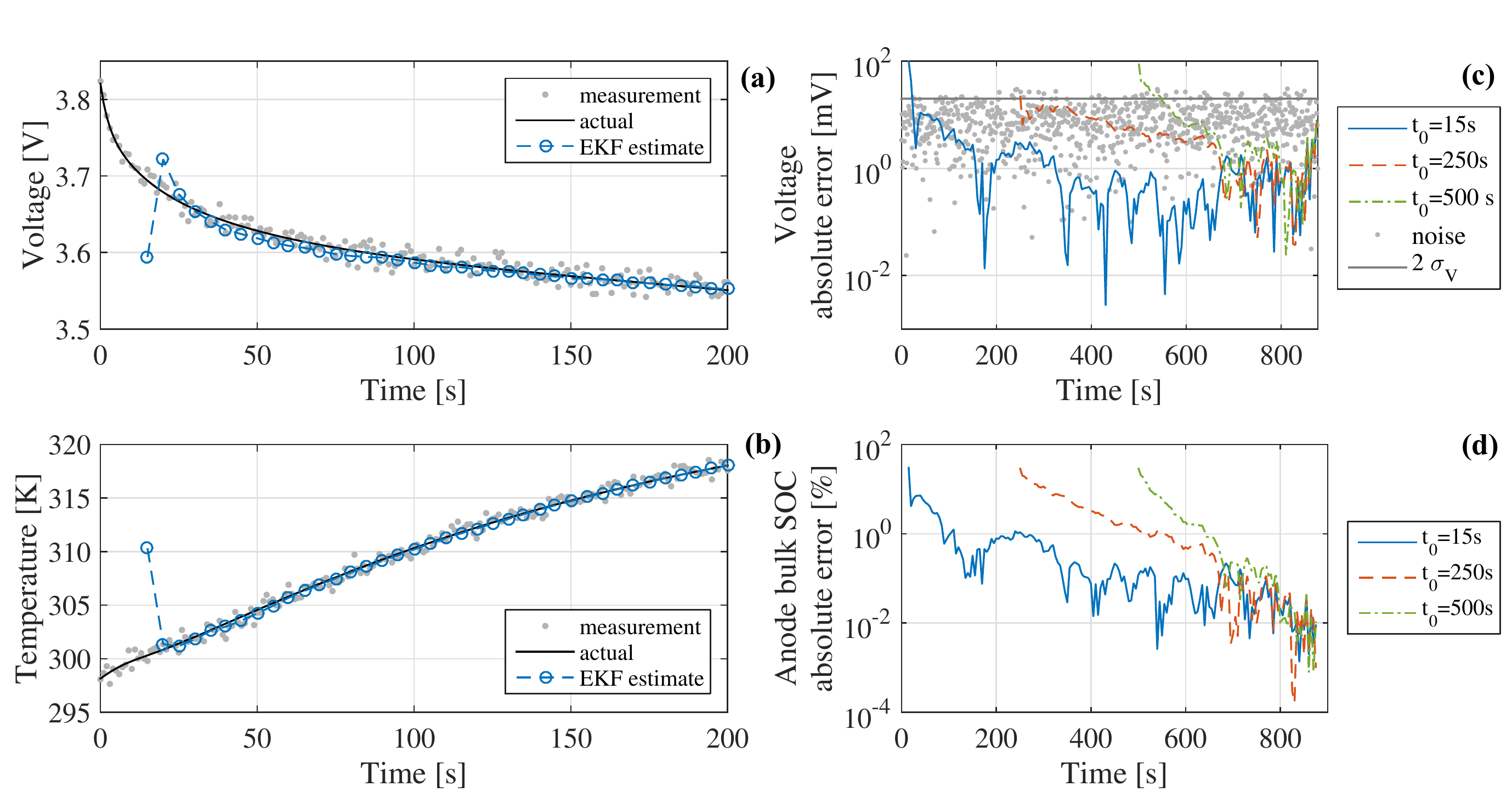}
\caption{Evolution of the voltage (a) and temperature (b) computed by the EKF compared to the actual and noisy measurements during the first 200~s of a 4C constant-current discharge.
Evolution of the absolute error on voltage (c) and anode bulk SOC (d) estimated by the EKF compared to actual values generated by the reference simulation during the full 4C constant-current discharge.}
\label{fig:Figure7}
\end{figure}

As a reference case, numerical experiments were computed by integrating the thermal-electrochemical model from 100~\% SOC until the 2~V minimum cut-off voltage under constant-current discharge and the CADC discussed in Section~\ref{sec:modelResultsAndDiscussion}.
The EKF was then started from several initial conditions to check its convergence behaviour at different SOC ranging from 100~\% to 50~\%.
For both the CADC and constant-current tests, the EKF initial guess on states assumed a cell at equilibrium (i.e. no concentration gradients) with an error on both the anode and cathode SOC of 30~\%. The initial error on the temperature was set to 10\textdegree C.
The variances for the generation of the additive Gaussian white measurement noise were set to $\sigma_V^2 = 1 \times 10^{-4}~V^2$ and $\sigma_T^2 = 0.25~K^2$ for the voltage and temperature respectively. These variances correspond to a standard deviation $\sigma_V = 10~mV$ on the voltage and $\sigma_T = 0.5~K$ on the temperature.
The measurement noise covariance matrix $R$ of the EKF was defined using these values.
No process noise was added to the state variables of the model and therefore the EKF process noise covariance matrix $Q$ was set to zero.

Fig.~\ref{fig:Figure7} shows the voltage and temperature calculated by the EKF and the corresponding measurements for the first 200~s of a 4C constant-current discharge with a 5~s time step. The EKF voltage and temperature rapidly converge to the actual reference values in only a few time steps.
The voltage absolute error for the full 4C discharge is shown on Fig.~\ref{fig:Figure7}c. The EKF was started at three different initial times, 15~s, 250~s and  500~s, to check the convergence behaviour at different SOCs. The grey line represents the 95~\% confidence interval on the voltage noisy measurement ($2 \sigma_V$) and the grey dots are the absolute measurement error on voltage. The EKF shows similar convergence behaviour for all initial time $t_0$ studied and the voltage estimate falls below the 95~\% confidence interval within the first few time steps. Similar results were observed for temperature measurements.

\begin{figure}[t]
\centering
\includegraphics[width=0.95\textwidth]{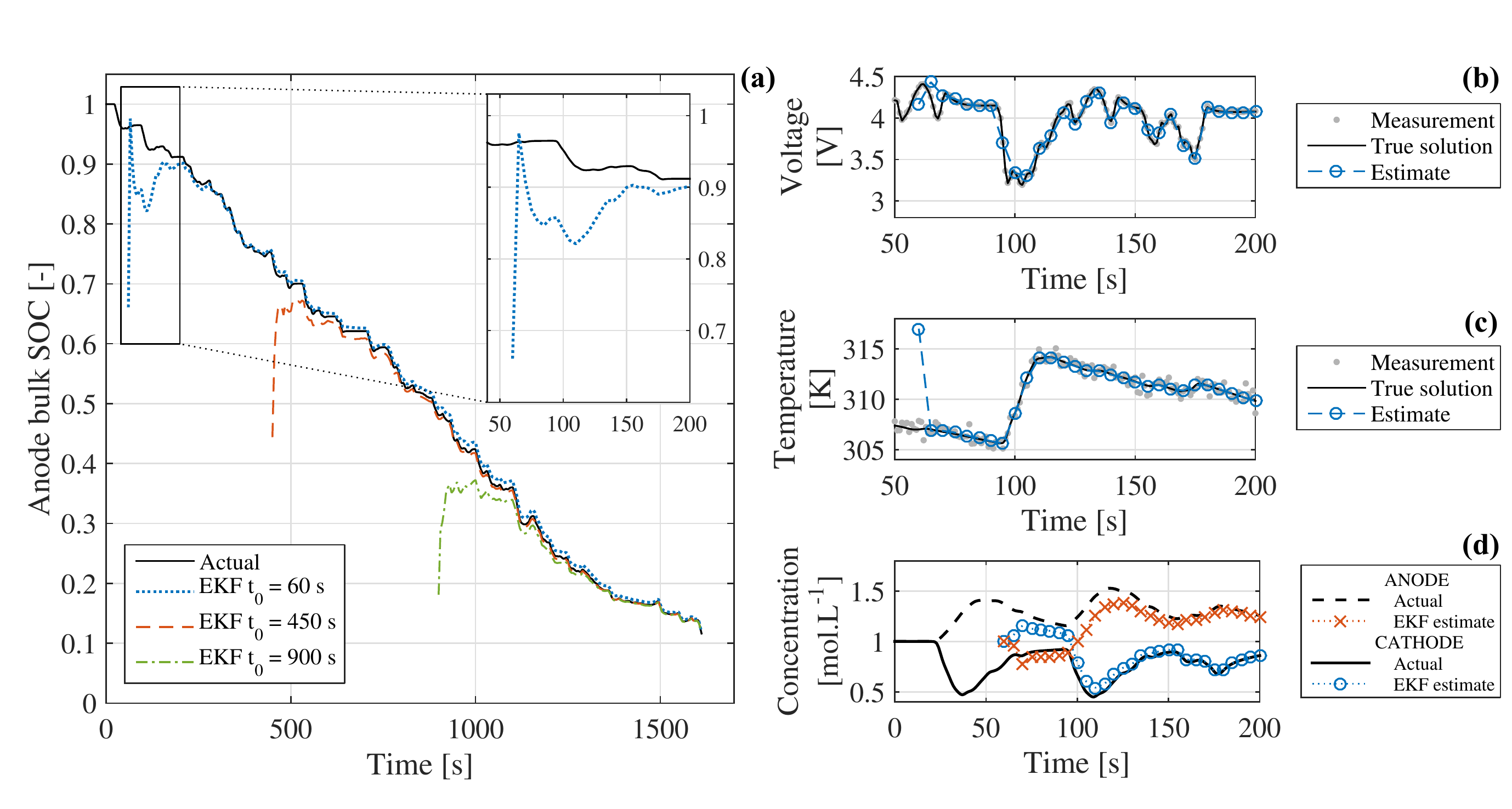}
\caption{Evolution of the anode bulk SOC estimated by the EKF compared to the actual value for CADC charge/discharge cycle (a). Evolution of the voltage (b), temperature (c) and electrolyte lithium concentration at the anode and cathode current collectors (d) computed by the EKF compared to the actual and noisy values generated by the reference simulation during the first 200~s of the CADC charge/discharge cycle.}
\label{fig:Figure8}
\end{figure}

The EKF algorithm is designed to accurately fit the measurements but this does not guarantee the convergence of the state estimates.
Fig.~\ref{fig:Figure7}d shows the absolute error between the EKF estimated anode bulk SOC and the actual SOC for the full 4C constant-current discharge.
Again, the EKF was started at different SOCs (initial times $t_0$) and shows a satisfactory convergence behaviour for all cases.
From the 30~\% initial absolute error, the anode bulk SOC estimate error falls below 1~\% after less than 200~s of simulation (300~s when started at $t_0 = 250~s$). Similar results were observed for the cathode bulk SOC due to the conservation of lithium in the cell.

The EKF algorithm was then applied to the battery charge/discharge cycle under the CADC. This drive cycle is highly dynamic with large current peaks. However, the EKF was solved using a 5~s time-step to reduce the computation time. The EKF under CADC was solved in 0.6~s of computation per second of simulation on average on the desktop computer previously mentioned. Fig.~\ref{fig:Figure8}b and Fig.~\ref{fig:Figure8}c show the rapid convergence of both the EKF voltage and temperature compared to the measurements.
Similarly to the constant-current scenario, the anode bulk SOC also converges relatively quickly to the actual value. The absolute error on anode bulk SOC falls below 1~\% by 150~s of simulation.

Another interesting state of the model is electrolyte concentration, since saturation or depletion of lithium in the electrolyte can lead to battery performance limitations at high current peaks. Fig.~\ref{fig:Figure8}d shows the evolution of the estimated and actual electrolyte concentration at the anode and cathode current collectors during the first few seconds of the driving cycle. This graphs shows that the algorithm is able to recover from the wrong initial conditions and track the electrolyte concentration during battery operation accurately.

\section{Conclusion}
A physics-based thermal-electrochemical spatially-distributed pseudo-2D model for lithium-ion batteries, so-called Newman model in the literature, is solved with Chebyshev orthogonal collocation in MATLAB. This results in a highly reduced number of states and computation cost compared to commonly employed finite-difference or finite-elements methods while maintaining accuracy. Comparative results against a much higher-order model solved in COMSOL Multiphysics confirm that accuracy  is preserved up to high C-rates. The relatively low number of states required with this approach enables our implementation of the model to be combined to a state observer for the estimation of battery internal states.

We used the extended Kalman filter to estimate the states of the pseudo-2D battery model due to its relatively low computational cost compared to other observers for non-linear models. The extended Kalman filter is able to estimate the state error by using a time-varying linear approximation of the model differential-algebraic equations about the state estimate at every time-step. To our knowledge, this work is the first attempt at estimating the internal states of the fully spatially-distributed pseudo-2D battery model using an extended Kalman filter. Results indicates that our state estimation algorithm is able to quickly (less than 200~s) recover the states of the model even with a 30~\% error on the initial SOC. This approach could be used within advanced battery management embedded systems for accurate battery state estimation and coupled to degradation models for health-conscious battery control.
Further work is being undertaken on investigating the observability and identifiability of the model states and parameters, and developing a parameter estimation algorithm.
%
%
%
%
%
\section*{Acknowledgements}
This work is funded by Samsung Electronics Co. Ltd. through a Global Research Outreach program in collaboration with the Samsung Advanced Institute of Technology. Shi Zhao is funded through the RCUK Energy Programme's STABLE-NET project (ref.\ EP/L014343/1).

\begin{appendices}
\section{Change of variable for the spherical particle diffusion model}
\label{app:changeOfVar}
The spherical diffusion model for the solid-phase particles is modified using the change of variable $\bar{c}_s = r c_s$. Assuming constant diffusivity, the diffusion equation becomes:
\begin{equation}
 \label{eq:P2Dmodel_solidPhase_diffusion_varChange}
\frac{\partial \bar{c}_{s}}{\partial t} = D_{s} \frac{\partial^2 \bar{c}_{s}}{\partial r^2}
\end{equation}
with the following boundary conditions:
\begin{align}
\label{eq:P2Dmodel_solidPhase_diffusion_varChange_bc_Rs}
\left. \frac{\partial \bar{c}_{s}}{\partial r} \right|_{r = R_{s}} - \frac{\bar{c}_{s}(r  = R_{s})}{R_{s}} &=  \frac{-R_{s}}{a_{s} \mathcal{F} D_{s}} j^{Li}\\
\bar{c}_s\left(r = 0 \right) &= 0 \label{eq:P2Dmodel_solidPhase_diffusion_varChange_bc_0}
\end{align}

\section{Interface boundary conditions for the domain decomposition}
\label{app:interfaceBC_domainDecomposition}
\subsection*{Continuity of electrolyte electric potential}
\begin{align}
\phi_{e,a}(x=L_{a}) &= \phi_{e,s}(x=L_{a}) \label{eq:P2Dmodel_electrolyte_potential_bc_L12} \\ 
\phi_{e,s}(x=L_{c}) &= \phi_{e,c}(x=L_{c})  \label{eq:P2Dmodel_electrolyte_potential_bc_L23}
\end{align}
\subsection*{Continuity of the concentration profile}
\begin{align}
c_{e,a}(x=L_{a}) &= c_{e,s}(x=L_{a}) \label{eq:P2Dmodel_electrolyte_diffusion_bc_L12_C0} \\ 
c_{e,s}(x=L_{c}) &= c_{e,c}(x=L_{c})  \label{eq:P2Dmodel_electrolyte_diffusion_bc_L23_C0}
\end{align}
\subsection*{Continuity of lithium ion flux}
\begin{align}
D^{eff}_{e,a} (L_a) \left. \frac{\partial c_{e,a}}{\partial x} \right|_{x = L_{a}} &= D^{eff}_{e,s} (L_a) \left. \frac{\partial c_{e,s}}{\partial x} \right|_{x = L_{a}}  \label{eq:P2Dmodel_electrolyte_diffusion_bc_L12_flux}\\
D^{eff}_{e,s} (L_c) \left. \frac{\partial c_{e,s}}{\partial x} \right|_{x = L_{c}} &= D^{eff}_{e,c} (L_c) \left. \frac{\partial c_{e,c}}{\partial x} \right|_{x = L_{c}} \label{eq:P2Dmodel_electrolyte_diffusion_bc_L23_flux}
\end{align}

\section{COMSOL implementation of the thermal-electrochemical P2D model}
\label{app:comsolImplementation}
The COMSOL implementation of the thermal-electrochemical P2D model discussed in Section~\ref{subsec:electrochemicalModel} is similar to \cite{Cai2011a} and involves using the COMSOL \emph{PDE Interfaces} and the \emph{ODE and DAE Interfaces} for equation-based modelling.
The macro-scale cell model is described on a 1D geometry divided into three regions (anode, separator and cathode) in the \textit{x}-direction. The micro-scale particle model is described on a 2D geometry in which the diffusion coefficients and electronic conductivity are set to zero in the \textit{x}-direction \cite{Cai2011a}. This reduces the 2D geometry to a 1D geometry in the \textit{y}-direction distributed along the cell thickness (\textit{x}-direction) that is equivalent to the radial \textit{r}-direction of the solid-phase spherical particles of the P2D model. The equations solved on the 1D and 2D geometries are coupled by projecting the local reaction rate $j^{Li}$ and the solid-phase surface concentration $c_s^{surf}$ from one geometry to the other using the \emph{linear extrusion} COMSOL function.
The 1D geometry was discretised using a uniformly spaced mesh with 22~elements, 8~elements and 21~elements in the anode, separator and cathode domains respectively. The 2D geometries of the electrodes were discretised using triangular elements for the core of the particles and quadrilateral elements at the surface of the particles. The anode was discretised with 890~triangular elements and 528~quadrilateral elements and the cathode with 588~triangular elements and 504~quadrilateral elements. This results in a COMSOL finite-element model with 7,856~degrees of freedom.
\end{appendices}

\section*{References}
\bibliographystyle{elsarticle-num-names} 
\bibliography{main}

\begin{thebibliography}{51}
\providecommand{\natexlab}[1]{#1}
\providecommand{\url}[1]{\texttt{#1}}
\providecommand{\urlprefix}{URL }
\expandafter\ifx\csname urlstyle\endcsname\relax
  \providecommand{\doi}[1]{doi:\discretionary{}{}{}#1}\else
  \providecommand{\doi}[1]{doi:\discretionary{}{}{}\begingroup
  \urlstyle{rm}\url{#1}\endgroup}\fi
\providecommand{\bibinfo}[2]{#2}

\bibitem[{Howey and Alavi(2015)}]{Howey2015}
\bibinfo{author}{D.~A. Howey}, \bibinfo{author}{S.~M. Alavi},
  \bibinfo{title}{{Rechargeable battery energy storage system design}}, in:
  \bibinfo{booktitle}{Handb. Clean Energy Syst. vol. 5},
  \bibinfo{publisher}{Wiley}, \bibinfo{year}{2015}.

\bibitem[{Plett(2004)}]{Plett2004d}
\bibinfo{author}{G.~L. Plett}, \bibinfo{title}{{Extended Kalman filtering for
  battery management systems of LiPB-based HEV battery packs - Part 2. Modeling
  and identification}}, \bibinfo{journal}{J. Power Sources}
  \bibinfo{volume}{134} (\bibinfo{year}{2004}) \bibinfo{pages}{262--276}.

\bibitem[{Hu et~al.(2012)Hu, Li, and Peng}]{Hu2012}
\bibinfo{author}{X.~Hu}, \bibinfo{author}{S.~Li}, \bibinfo{author}{H.~Peng},
  \bibinfo{title}{{A comparative study of equivalent circuit models for Li-ion
  batteries}}, \bibinfo{journal}{J. Power Sources} \bibinfo{volume}{198}
  (\bibinfo{year}{2012}) \bibinfo{pages}{359--367}, ISSN
  \bibinfo{issn}{03787753},
  \doi{\bibinfo{doi}{10.1016/j.jpowsour.2011.10.013}}.

\bibitem[{Birkl and Howey(2013)}]{Birkl2013}
\bibinfo{author}{C.~Birkl}, \bibinfo{author}{D.~Howey}, \bibinfo{title}{{Model
  identification and parameter estimation for LiFePO4 batteries}}, in:
  \bibinfo{booktitle}{Hybrid Electr. Veh. Conf. 2013 (HEVC 2013)}, ISBN
  \bibinfo{isbn}{978-1-84919-776-2}, \bibinfo{pages}{2.1--2.1},
  \doi{\bibinfo{doi}{10.1049/cp.2013.1889}}, \bibinfo{year}{2013}.

\bibitem[{Doyle et~al.(1993)Doyle, Fuller, and Newman}]{Doyle1993}
\bibinfo{author}{M.~Doyle}, \bibinfo{author}{T.~F. Fuller},
  \bibinfo{author}{J.~Newman}, \bibinfo{title}{{Modeling of Galvanostatic
  Charge and Discharge of the Lithium/Polymer/Insertion Cell}},
  \bibinfo{journal}{J. Electrochem. Soc.}
  \bibinfo{volume}{140}~(\bibinfo{number}{6}) (\bibinfo{year}{1993})
  \bibinfo{pages}{1526}, ISSN \bibinfo{issn}{00134651},
  \doi{\bibinfo{doi}{10.1149/1.2221597}}.

\bibitem[{Doyle(1995)}]{Doyle1995}
\bibinfo{author}{M.~Doyle}, \bibinfo{title}{{Design and simulation of lithium
  rechargeable batteries}}, Ph.D. thesis, \bibinfo{school}{University of
  California, Berkeley Laboratory}, \bibinfo{year}{1995}.

\bibitem[{Fuller et~al.(1994)Fuller, Doyle, and Newman}]{Fuller1994}
\bibinfo{author}{T.~F. Fuller}, \bibinfo{author}{M.~Doyle},
  \bibinfo{author}{J.~Newman}, \bibinfo{title}{{Simulation and Optimization of
  the Dual Lithium Ion Insertion Cell}}, \bibinfo{journal}{J. Electrochem.
  Soc.} \bibinfo{volume}{141}~(\bibinfo{number}{1}) (\bibinfo{year}{1994})
  \bibinfo{pages}{1--10}.

\bibitem[{Doyle and Fuentes(2003)}]{Doyle2003}
\bibinfo{author}{M.~Doyle}, \bibinfo{author}{Y.~Fuentes},
  \bibinfo{title}{{Computer Simulations of a Lithium-Ion Polymer Battery and
  Implications for Higher Capacity Next-Generation Battery Designs}},
  \bibinfo{journal}{J. Electrochem. Soc.}
  \bibinfo{volume}{150}~(\bibinfo{number}{6}) (\bibinfo{year}{2003})
  \bibinfo{pages}{A706}, ISSN \bibinfo{issn}{00134651},
  \doi{\bibinfo{doi}{10.1149/1.1569478}}.

\bibitem[{Forman et~al.(2012)Forman, Moura, Stein, and Fathy}]{Forman2012}
\bibinfo{author}{J.~C. Forman}, \bibinfo{author}{S.~J. Moura},
  \bibinfo{author}{J.~L. Stein}, \bibinfo{author}{H.~K. Fathy},
  \bibinfo{title}{{Genetic identification and fisher identifiability analysis
  of the Doyle-Fuller-Newman model from experimental cycling of a LiFePO4
  cell}}, \bibinfo{journal}{J. Power Sources} \bibinfo{volume}{210}
  (\bibinfo{year}{2012}) \bibinfo{pages}{263--275}, ISSN
  \bibinfo{issn}{03787753},
  \doi{\bibinfo{doi}{10.1016/j.jpowsour.2012.03.009}}.

\bibitem[{Santhanagopalan and White(2006)}]{Santhanagopalan2006a}
\bibinfo{author}{S.~Santhanagopalan}, \bibinfo{author}{R.~E. White},
  \bibinfo{title}{{Online estimation of the state of charge of a lithium ion
  cell}}, \bibinfo{journal}{J. Power Sources}
  \bibinfo{volume}{161}~(\bibinfo{number}{2}) (\bibinfo{year}{2006})
  \bibinfo{pages}{1346--1355}, ISSN \bibinfo{issn}{03787753},
  \doi{\bibinfo{doi}{10.1016/j.jpowsour.2006.04.146}}.

\bibitem[{Moura et~al.(2013{\natexlab{a}})Moura, Chaturvedi, and
  Krsti\'{c}}]{Moura2013b}
\bibinfo{author}{S.~J. Moura}, \bibinfo{author}{N.~A. Chaturvedi},
  \bibinfo{author}{M.~Krsti\'{c}}, \bibinfo{title}{{Adaptive Partial
  Differential Equation Observer for Battery State-of-Charge/State-of-Health
  Estimation Via an Electrochemical Model}}, \bibinfo{journal}{J. Dyn. Syst.
  Meas. Control} \bibinfo{volume}{136}~(\bibinfo{number}{1})
  (\bibinfo{year}{2013}{\natexlab{a}}) \bibinfo{pages}{011015}, ISSN
  \bibinfo{issn}{0022-0434}, \doi{\bibinfo{doi}{10.1115/1.4024801}}.

\bibitem[{{Di Domenico} et~al.(2010){Di Domenico}, Stefanopoulou, and
  Fiengo}]{DiDomenico2010}
\bibinfo{author}{D.~{Di Domenico}}, \bibinfo{author}{A.~Stefanopoulou},
  \bibinfo{author}{G.~Fiengo}, \bibinfo{title}{{Lithium-Ion Battery State of
  Charge and Critical Surface Charge Estimation Using an Electrochemical
  Model-Based Extended Kalman Filter}}, \bibinfo{journal}{J. Dyn. Syst. Meas.
  Control} \bibinfo{volume}{132}~(\bibinfo{number}{6}) (\bibinfo{year}{2010})
  \bibinfo{pages}{061302}, ISSN \bibinfo{issn}{00220434},
  \doi{\bibinfo{doi}{10.1115/1.4002475}}.

\bibitem[{Smith et~al.(2008{\natexlab{a}})Smith, Rahn, and Wang}]{Smith2008b}
\bibinfo{author}{K.~Smith}, \bibinfo{author}{C.~D. Rahn},
  \bibinfo{author}{C.-Y. Wang}, \bibinfo{title}{{Model-based electrochemical
  estimation of lithium-ion batteries}}, \bibinfo{journal}{2008 IEEE Int. Conf.
  Control Appl.} ~(\bibinfo{number}{1}) (\bibinfo{year}{2008}{\natexlab{a}})
  \bibinfo{pages}{714--719}, \doi{\bibinfo{doi}{10.1109/CCA.2008.4629589}}.

\bibitem[{Smith et~al.(2010)Smith, Rahn, and Wang}]{Smith2010}
\bibinfo{author}{K.~Smith}, \bibinfo{author}{C.~D. Rahn},
  \bibinfo{author}{C.-Y. Wang}, \bibinfo{title}{{Model-Based Electrochemical
  Estimation and Constraint Management for Pulse Operation of Lithium Ion
  Batteries}}, \bibinfo{journal}{IEEE Trans. Control Syst. Technol.}
  \bibinfo{volume}{18}~(\bibinfo{number}{3}) (\bibinfo{year}{2010})
  \bibinfo{pages}{654--663}, ISSN \bibinfo{issn}{1063-6536},
  \doi{\bibinfo{doi}{10.1109/TCST.2009.2027023}}.

\bibitem[{Smith et~al.(2007)Smith, Rahn, and Wang}]{Smith2007b}
\bibinfo{author}{K.~Smith}, \bibinfo{author}{C.~D. Rahn},
  \bibinfo{author}{C.-Y. Wang}, \bibinfo{title}{{Control oriented 1D
  electrochemical model of lithium ion battery}}, \bibinfo{journal}{Energy
  Convers. Manag.} \bibinfo{volume}{48}~(\bibinfo{number}{9})
  (\bibinfo{year}{2007}) \bibinfo{pages}{2565--2578}, ISSN
  \bibinfo{issn}{01968904},
  \doi{\bibinfo{doi}{10.1016/j.enconman.2007.03.015}}.

\bibitem[{Smith et~al.(2008{\natexlab{b}})Smith, Rahn, and Wang}]{Smith2008}
\bibinfo{author}{K.~Smith}, \bibinfo{author}{C.~D. Rahn},
  \bibinfo{author}{C.-Y. Wang}, \bibinfo{title}{{Model Order Reduction of 1D
  Diffusion Systems Via Residue Grouping}}, \bibinfo{journal}{J. Dyn. Syst.
  Meas. Control} \bibinfo{volume}{130}~(\bibinfo{number}{1})
  (\bibinfo{year}{2008}{\natexlab{b}}) \bibinfo{pages}{011012}, ISSN
  \bibinfo{issn}{00220434}, \doi{\bibinfo{doi}{10.1115/1.2807068}}.

\bibitem[{Stetzel et~al.(2015)Stetzel, Aldrich, Trimboli, and
  Plett}]{Stetzel2015a}
\bibinfo{author}{K.~D. Stetzel}, \bibinfo{author}{L.~L. Aldrich},
  \bibinfo{author}{M.~S. Trimboli}, \bibinfo{author}{G.~L. Plett},
  \bibinfo{title}{{Electrochemical state and internal variables estimation
  using a reduced-order physics-based model of a lithium-ion cell and an
  extended Kalman filter}}, \bibinfo{journal}{J. Power Sources}
  \bibinfo{volume}{278} (\bibinfo{year}{2015}) \bibinfo{pages}{490--505}, ISSN
  \bibinfo{issn}{03787753},
  \doi{\bibinfo{doi}{10.1016/j.jpowsour.2014.11.135}}.

\bibitem[{Lee et~al.(2012{\natexlab{a}})Lee, Chemistruck, and Plett}]{Lee2012a}
\bibinfo{author}{J.~L. Lee}, \bibinfo{author}{A.~Chemistruck},
  \bibinfo{author}{G.~L. Plett}, \bibinfo{title}{{One-dimensional physics-based
  reduced-order model of lithium-ion dynamics}}, \bibinfo{journal}{J. Power
  Sources} \bibinfo{volume}{220} (\bibinfo{year}{2012}{\natexlab{a}})
  \bibinfo{pages}{430--448}, ISSN \bibinfo{issn}{03787753},
  \doi{\bibinfo{doi}{10.1016/j.jpowsour.2012.07.075}}.

\bibitem[{Lee et~al.(2012{\natexlab{b}})Lee, Chemistruck, and Plett}]{Lee2012b}
\bibinfo{author}{J.~L. Lee}, \bibinfo{author}{A.~Chemistruck},
  \bibinfo{author}{G.~L. Plett}, \bibinfo{title}{{Discrete-time realization of
  transcendental impedance models, with application to modeling spherical solid
  diffusion}}, \bibinfo{journal}{J. Power Sources} \bibinfo{volume}{206}
  (\bibinfo{year}{2012}{\natexlab{b}}) \bibinfo{pages}{367--377}, ISSN
  \bibinfo{issn}{03787753},
  \doi{\bibinfo{doi}{10.1016/j.jpowsour.2012.01.134}}.

\bibitem[{Lee et~al.(2014)Lee, Aldrich, Stetzel, and Plett}]{Lee2014a}
\bibinfo{author}{J.~L. Lee}, \bibinfo{author}{L.~L. Aldrich},
  \bibinfo{author}{K.~D. Stetzel}, \bibinfo{author}{G.~L. Plett},
  \bibinfo{title}{{Extended operating range for reduced-order model of
  lithium-ion cells}}, \bibinfo{journal}{J. Power Sources}
  \bibinfo{volume}{255} (\bibinfo{year}{2014}) \bibinfo{pages}{85--100}, ISSN
  \bibinfo{issn}{03787753},
  \doi{\bibinfo{doi}{10.1016/j.jpowsour.2013.12.134}}.

\bibitem[{Dao et~al.(2012)Dao, Vyasarayani, and McPhee}]{Dao2012a}
\bibinfo{author}{T.-S. Dao}, \bibinfo{author}{C.~P. Vyasarayani},
  \bibinfo{author}{J.~McPhee}, \bibinfo{title}{{Simplification and order
  reduction of lithium-ion battery model based on porous-electrode theory}},
  \bibinfo{journal}{J. Power Sources} \bibinfo{volume}{198}
  (\bibinfo{year}{2012}) \bibinfo{pages}{329--337}, ISSN
  \bibinfo{issn}{03787753},
  \doi{\bibinfo{doi}{10.1016/j.jpowsour.2011.09.034}}.

\bibitem[{Cai and White(2012)}]{Cai2012}
\bibinfo{author}{L.~Cai}, \bibinfo{author}{R.~E. White},
  \bibinfo{title}{{Lithium ion cell modeling using orthogonal collocation on
  finite elements}}, \bibinfo{journal}{J. Power Sources} \bibinfo{volume}{217}
  (\bibinfo{year}{2012}) \bibinfo{pages}{248--255}, ISSN
  \bibinfo{issn}{03787753},
  \doi{\bibinfo{doi}{10.1016/j.jpowsour.2012.06.043}}.

\bibitem[{Northrop et~al.(2011)Northrop, Ramadesigan, De, and
  Subramanian}]{Northrop2011b}
\bibinfo{author}{P.~W.~C. Northrop}, \bibinfo{author}{V.~Ramadesigan},
  \bibinfo{author}{S.~De}, \bibinfo{author}{V.~R. Subramanian},
  \bibinfo{title}{{Coordinate Transformation, Orthogonal Collocation, Model
  Reformulation and Simulation of Electrochemical-Thermal Behavior of
  Lithium-Ion Battery Stacks}}, \bibinfo{journal}{J. Electrochem. Soc.}
  \bibinfo{volume}{158}~(\bibinfo{number}{12}) (\bibinfo{year}{2011})
  \bibinfo{pages}{A1461}, ISSN \bibinfo{issn}{00134651},
  \doi{\bibinfo{doi}{10.1149/2.058112jes}}.

\bibitem[{Suthar et~al.(2014)Suthar, Northrop, Braatz, and
  Subramanian}]{Suthar2014}
\bibinfo{author}{B.~Suthar}, \bibinfo{author}{P.~W.~C. Northrop},
  \bibinfo{author}{R.~D. Braatz}, \bibinfo{author}{V.~R. Subramanian},
  \bibinfo{title}{{Optimal Charging Profiles with Minimal Intercalation-Induced
  Stresses for Lithium-Ion Batteries Using Reformulated Pseudo 2-Dimensional
  Models}}, \bibinfo{journal}{J. Electrochem. Soc.}
  \bibinfo{volume}{161}~(\bibinfo{number}{11}) (\bibinfo{year}{2014})
  \bibinfo{pages}{F3144--F3155}, ISSN \bibinfo{issn}{0013-4651},
  \doi{\bibinfo{doi}{10.1149/2.0211411jes}}.

\bibitem[{Bizeray et~al.(2013)Bizeray, Duncan, and Howey}]{Bizeray2013}
\bibinfo{author}{A.~Bizeray}, \bibinfo{author}{S.~Duncan},
  \bibinfo{author}{D.~Howey}, \bibinfo{title}{{Advanced battery management
  systems using fast electrochemical modelling}}, in:
  \bibinfo{booktitle}{Hybrid Electr. Veh. Conf. 2013 (HEVC 2013)},
  \bibinfo{publisher}{Institution of Engineering and Technology}, ISBN
  \bibinfo{isbn}{978-1-84919-776-2}, \bibinfo{pages}{2.2--2.2},
  \doi{\bibinfo{doi}{10.1049/cp.2013.1890}}, \bibinfo{year}{2013}.

\bibitem[{Drummond et~al.(2015)Drummond, Howey, and Duncan}]{Drummond2015}
\bibinfo{author}{R.~Drummond}, \bibinfo{author}{D.~A. Howey},
  \bibinfo{author}{S.~R. Duncan}, \bibinfo{title}{{Low-Order Mathematical
  Modelling of Electric Double Layer Supercapacitors Using Spectral Methods}},
  \bibinfo{journal}{J. Power Sources} \bibinfo{volume}{277}
  (\bibinfo{year}{2015}) \bibinfo{pages}{317--328}.

\bibitem[{Suthar et~al.(2013)Suthar, Ramadesigan, Northrop, Gopaluni,
  Santhanagopalan, Braatz, and Subramanian}]{Suthar2013}
\bibinfo{author}{B.~Suthar}, \bibinfo{author}{V.~Ramadesigan},
  \bibinfo{author}{P.~W.~C. Northrop}, \bibinfo{author}{B.~Gopaluni},
  \bibinfo{author}{S.~Santhanagopalan}, \bibinfo{author}{R.~D. Braatz},
  \bibinfo{author}{V.~R. Subramanian}, \bibinfo{title}{{Optimal Control and
  State Estimation of Lithium-ion Batteries Using Reformulated Models}},
  \bibinfo{journal}{Am. Control Conf. (ACC), 2013}  (\bibinfo{year}{2013})
  \bibinfo{pages}{5350--5355}.

\bibitem[{Gopaluni and Braatz(2013)}]{Gopaluni2013}
\bibinfo{author}{R.~B. Gopaluni}, \bibinfo{author}{R.~D. Braatz},
  \bibinfo{title}{{State of Charge Estimation in Li-Ion Batteries Using an
  Isothermal Pseudo Two-Dimensional Model}}, \bibinfo{journal}{Proc. 10th IFAC
  Int. Symp. Dyn. Control Process Syst.}  (\bibinfo{year}{2013})
  \bibinfo{pages}{135--140}\doi{\bibinfo{doi}{10.3182/20131218-3-IN-2045.00163}}.

\bibitem[{Chaturvedi et~al.(2010)Chaturvedi, Klein, Christensen, Ahmed, and
  Kojic}]{Chaturvedi2010}
\bibinfo{author}{N.~Chaturvedi}, \bibinfo{author}{R.~Klein},
  \bibinfo{author}{J.~Christensen}, \bibinfo{author}{J.~Ahmed},
  \bibinfo{author}{A.~Kojic}, \bibinfo{title}{{Algorithms for Advanced
  Battery-Management Systems}}, \bibinfo{journal}{IEEE Control Syst. Mag.}
  \bibinfo{volume}{30}~(\bibinfo{number}{3}) (\bibinfo{year}{2010})
  \bibinfo{pages}{49--68}, ISSN \bibinfo{issn}{0272-1708},
  \doi{\bibinfo{doi}{10.1109/MCS.2010.936293}}.

\bibitem[{Moura et~al.(2013{\natexlab{b}})Moura, Stein, and Fathy}]{Moura2013}
\bibinfo{author}{S.~J. Moura}, \bibinfo{author}{J.~L. Stein},
  \bibinfo{author}{H.~K. Fathy}, \bibinfo{title}{{Battery-Health Conscious
  Power Management in Plug-In Hybrid Electric Vehicles via Electrochemical
  Modeling and Stochastic Control}}, \bibinfo{journal}{IEEE Trans. Control
  Syst. Technol.} \bibinfo{volume}{21}~(\bibinfo{number}{3})
  (\bibinfo{year}{2013}{\natexlab{b}}) \bibinfo{pages}{679--694}.

\bibitem[{Newman and Tiedemann(1975)}]{Newman1975}
\bibinfo{author}{J.~Newman}, \bibinfo{author}{W.~Tiedemann},
  \bibinfo{title}{{Porous-electrode theory with battery applications}},
  \bibinfo{journal}{AIChE J.} \bibinfo{volume}{21}~(\bibinfo{number}{1})
  (\bibinfo{year}{1975}) \bibinfo{pages}{25--41}, ISSN
  \bibinfo{issn}{0001-1541}, \doi{\bibinfo{doi}{10.1002/aic.690210103}}.

\bibitem[{Chung et~al.(2013)Chung, Ebner, Ely, Wood, and {Edwin
  Garc\'{\i}a}}]{Chung2013}
\bibinfo{author}{D.-W. Chung}, \bibinfo{author}{M.~Ebner},
  \bibinfo{author}{D.~R. Ely}, \bibinfo{author}{V.~Wood},
  \bibinfo{author}{R.~{Edwin Garc\'{\i}a}}, \bibinfo{title}{{Validity of the
  Bruggeman relation for porous electrodes}}, \bibinfo{journal}{Model. Simul.
  Mater. Sci. Eng.} \bibinfo{volume}{21} (\bibinfo{year}{2013})
  \bibinfo{pages}{074009}, ISSN \bibinfo{issn}{0965-0393},
  \doi{\bibinfo{doi}{10.1088/0965-0393/21/7/074009}}.

\bibitem[{Ramadass et~al.(2004)Ramadass, Haran, Gomadam, White, and
  Popov}]{Ramadass2004}
\bibinfo{author}{P.~Ramadass}, \bibinfo{author}{B.~Haran},
  \bibinfo{author}{P.~M. Gomadam}, \bibinfo{author}{R.~White},
  \bibinfo{author}{B.~N. Popov}, \bibinfo{title}{{Development of First
  Principles Capacity Fade Model for Li-Ion Cells}}, \bibinfo{journal}{J.
  Electrochem. Soc.} \bibinfo{volume}{151}~(\bibinfo{number}{2})
  (\bibinfo{year}{2004}) \bibinfo{pages}{A196}, ISSN \bibinfo{issn}{00134651},
  \doi{\bibinfo{doi}{10.1149/1.1634273}}.

\bibitem[{Ramos and Please(2015)}]{Ramos2015}
\bibinfo{author}{A.~M. Ramos}, \bibinfo{author}{C.~P. Please},
  \bibinfo{title}{{Some comments on the Butler-Volmer equation for modeling
  Lithium-ion batteries}}, \bibinfo{journal}{ArXiv e-prints}
  (\bibinfo{year}{2015}) \bibinfo{pages}{1--14}.

\bibitem[{Smith and Wang(2006)}]{Smith2006b}
\bibinfo{author}{K.~Smith}, \bibinfo{author}{C.-Y. Wang},
  \bibinfo{title}{{Power and thermal characterization of a lithium-ion battery
  pack for hybrid-electric vehicles}}, \bibinfo{journal}{J. Power Sources}
  \bibinfo{volume}{160}~(\bibinfo{number}{1}) (\bibinfo{year}{2006})
  \bibinfo{pages}{662--673}, ISSN \bibinfo{issn}{03787753},
  \doi{\bibinfo{doi}{10.1016/j.jpowsour.2006.01.038}}.

\bibitem[{Valo̸en and Reimers(2005)}]{Valoen2005}
\bibinfo{author}{L.~O. Valo̸en}, \bibinfo{author}{J.~N. Reimers},
  \bibinfo{title}{{Transport Properties of LiPF6-Based Li-Ion Battery
  Electrolytes}}, \bibinfo{journal}{J. Electrochem. Soc.}
  \bibinfo{volume}{152}~(\bibinfo{number}{5}) (\bibinfo{year}{2005})
  \bibinfo{pages}{A882}, ISSN \bibinfo{issn}{00134651},
  \doi{\bibinfo{doi}{10.1149/1.1872737}}.

\bibitem[{Kumaresan et~al.(2008)Kumaresan, Sikha, and White}]{Kumaresan2008}
\bibinfo{author}{K.~Kumaresan}, \bibinfo{author}{G.~Sikha},
  \bibinfo{author}{R.~E. White}, \bibinfo{title}{{Thermal Model for a Li-Ion
  Cell}}, \bibinfo{journal}{J. Electrochem. Soc.}
  \bibinfo{volume}{155}~(\bibinfo{number}{2}) (\bibinfo{year}{2008})
  \bibinfo{pages}{A164}, ISSN \bibinfo{issn}{00134651},
  \doi{\bibinfo{doi}{10.1149/1.2817888}}.

\bibitem[{Wu et~al.(2013)Wu, Yufit, Marinescu, Offer, Martinez-Botas, and
  Brandon}]{Wu2013c}
\bibinfo{author}{B.~Wu}, \bibinfo{author}{V.~Yufit},
  \bibinfo{author}{M.~Marinescu}, \bibinfo{author}{G.~J. Offer},
  \bibinfo{author}{R.~F. Martinez-Botas}, \bibinfo{author}{N.~P. Brandon},
  \bibinfo{title}{{Coupled thermal-electrochemical modelling of uneven heat
  generation in lithium-ion battery packs}}, \bibinfo{journal}{J. Power
  Sources} \bibinfo{volume}{243} (\bibinfo{year}{2013})
  \bibinfo{pages}{544--554}, ISSN \bibinfo{issn}{03787753},
  \doi{\bibinfo{doi}{10.1016/j.jpowsour.2013.05.164}}.

\bibitem[{Guo et~al.(2011)Guo, Sikha, and White}]{Guo2011}
\bibinfo{author}{M.~Guo}, \bibinfo{author}{G.~Sikha}, \bibinfo{author}{R.~E.
  White}, \bibinfo{title}{{Single-Particle Model for a Lithium-Ion Cell:
  Thermal Behavior}}, \bibinfo{journal}{J. Electrochem. Soc.}
  \bibinfo{volume}{158}~(\bibinfo{number}{2}) (\bibinfo{year}{2011})
  \bibinfo{pages}{A122}, ISSN \bibinfo{issn}{00134651},
  \doi{\bibinfo{doi}{10.1149/1.3521314}}.

\bibitem[{Egorkina and Skundin(1998)}]{Egorkina1998a}
\bibinfo{author}{O.~Y. Egorkina}, \bibinfo{author}{A.~M. Skundin},
  \bibinfo{title}{{The effect of temperature on lithium intercalation into
  carbon materials}}, \bibinfo{journal}{J. Solid State Electrochem.}
  \bibinfo{volume}{2}~(\bibinfo{number}{4}) (\bibinfo{year}{1998})
  \bibinfo{pages}{216--220}, ISSN \bibinfo{issn}{1432-8488},
  \doi{\bibinfo{doi}{10.1007/s100080050091}}.

\bibitem[{Kulova et~al.(2006)Kulova, Skundin, Nizhnikovskii, and
  Fesenko}]{Kulova2006}
\bibinfo{author}{T.~L. Kulova}, \bibinfo{author}{A.~M. Skundin},
  \bibinfo{author}{E.~A. Nizhnikovskii}, \bibinfo{author}{A.~V. Fesenko},
  \bibinfo{title}{{Temperature effect on the lithium diffusion rate in
  graphite}}, \bibinfo{journal}{Russ. J. Electrochem.}
  \bibinfo{volume}{42}~(\bibinfo{number}{3}) (\bibinfo{year}{2006})
  \bibinfo{pages}{259--262}, ISSN \bibinfo{issn}{1023-1935},
  \doi{\bibinfo{doi}{10.1134/S1023193506030086}}.

\bibitem[{Nakamura et~al.(2000)Nakamura, Ohno, Okamura, Michihiro, Nakabayashi,
  and Kanashiro}]{Nakamura2000}
\bibinfo{author}{K.~Nakamura}, \bibinfo{author}{H.~Ohno},
  \bibinfo{author}{K.~Okamura}, \bibinfo{author}{Y.~Michihiro},
  \bibinfo{author}{I.~Nakabayashi}, \bibinfo{author}{T.~Kanashiro},
  \bibinfo{title}{{On the diffusion of Li+ defects in LiCoO2 and LiNiO2}},
  \bibinfo{journal}{Solid State Ionics}
  \bibinfo{volume}{135}~(\bibinfo{number}{1-4}) (\bibinfo{year}{2000})
  \bibinfo{pages}{143--147}, ISSN \bibinfo{issn}{01672738},
  \doi{\bibinfo{doi}{10.1016/S0167-2738(00)00293-9}}.

\bibitem[{Zheng et~al.(2005)Zheng, Qin, Zhao, Abe, and Ogumi}]{Zheng2005}
\bibinfo{author}{H.~Zheng}, \bibinfo{author}{J.~Qin},
  \bibinfo{author}{Y.~Zhao}, \bibinfo{author}{T.~Abe},
  \bibinfo{author}{Z.~Ogumi}, \bibinfo{title}{{Temperature dependence of the
  electrochemical behavior of LiCoO in quaternary ammonium-based ionic liquid
  electrolyte}}, \bibinfo{journal}{Solid State Ionics}
  \bibinfo{volume}{176}~(\bibinfo{number}{29-30}) (\bibinfo{year}{2005})
  \bibinfo{pages}{2219--2226}, ISSN \bibinfo{issn}{01672738},
  \doi{\bibinfo{doi}{10.1016/j.ssi.2005.06.020}}.

\bibitem[{Shampine et~al.(1999)Shampine, Reichelt, and
  Kierzenka}]{Shampine1999}
\bibinfo{author}{L.~F. Shampine}, \bibinfo{author}{M.~W. Reichelt},
  \bibinfo{author}{J.~A. Kierzenka}, \bibinfo{title}{{Solving Index-1 DAEs in
  MATLAB and Simulink}}, \bibinfo{journal}{SIAM Rev.}
  \bibinfo{volume}{41}~(\bibinfo{number}{3}) (\bibinfo{year}{1999})
  \bibinfo{pages}{538--552}, ISSN \bibinfo{issn}{0036-1445},
  \doi{\bibinfo{doi}{10.1137/S003614459933425X}}.

\bibitem[{Trefethen(2000)}]{Trefethen2000}
\bibinfo{author}{L.~N. Trefethen}, \bibinfo{title}{{Spectral Methods in
  MATLAB}}, \bibinfo{publisher}{SIAM}, ISBN \bibinfo{isbn}{978-0-89871-465-4},
  \urlprefix\url{http://epubs.siam.org/doi/book/10.1137/1.9780898719598},
  \bibinfo{year}{2000}.

\bibitem[{Gottlieb and Orszag(1977)}]{Gottlieb1977}
\bibinfo{author}{D.~Gottlieb}, \bibinfo{author}{S.~A. Orszag},
  \bibinfo{title}{{Numerical analysis of spectral methods: Theory and
  Applications}}, \bibinfo{publisher}{SIAM}, ISBN \bibinfo{isbn}{0-89871-023-5
  (paperback)}, \doi{\bibinfo{doi}{10.1137/1.9781611970425}},
  \urlprefix\url{http://hdl.handle.net/2060/19790002644}, \bibinfo{year}{1977}.

\bibitem[{Weideman and Reddy(2000)}]{Weideman2000}
\bibinfo{author}{J.~A.~C. Weideman}, \bibinfo{author}{S.~C. Reddy},
  \bibinfo{title}{{A MATLAB Differentiation Matrix Suite}},
  \bibinfo{journal}{ACM Trans. Math. Softwares}
  \bibinfo{volume}{26}~(\bibinfo{number}{4}) (\bibinfo{year}{2000})
  \bibinfo{pages}{465--519}.

\bibitem[{Cai and White(2011)}]{Cai2011a}
\bibinfo{author}{L.~Cai}, \bibinfo{author}{R.~E. White},
  \bibinfo{title}{{Mathematical modeling of a lithium ion battery with thermal
  effects in COMSOL Inc. Multiphysics (MP) software}}, \bibinfo{journal}{J.
  Power Sources} \bibinfo{volume}{196}~(\bibinfo{number}{14})
  (\bibinfo{year}{2011}) \bibinfo{pages}{5985--5989}, ISSN
  \bibinfo{issn}{03787753},
  \doi{\bibinfo{doi}{10.1016/j.jpowsour.2011.03.017}}.

\bibitem[{Andr\'{e}(2004)}]{Andre2004}
\bibinfo{author}{M.~Andr\'{e}}, \bibinfo{title}{{The ARTEMIS European driving
  cycles for measuring car pollutant emissions}}, \bibinfo{journal}{Sci. Total
  Environ.} \bibinfo{volume}{334-335} (\bibinfo{year}{2004})
  \bibinfo{pages}{73--84}, ISSN \bibinfo{issn}{00489697},
  \doi{\bibinfo{doi}{10.1016/j.scitotenv.2004.04.070}}.

\bibitem[{Becerra et~al.(2001)Becerra, Roberts, and Griffiths}]{Becerra2001}
\bibinfo{author}{V.~M. Becerra}, \bibinfo{author}{P.~D. Roberts},
  \bibinfo{author}{G.~W. Griffiths}, \bibinfo{title}{{Applying the extended
  Kalman filter to systems described by nonlinear differential-algebraic
  equations}}, \bibinfo{journal}{Control Eng. Pract.} \bibinfo{volume}{9}
  (\bibinfo{year}{2001}) \bibinfo{pages}{267--281}.

\bibitem[{Gelb et~al.(1974)Gelb, Kasper, Nash, Price, and
  Sutherland}]{Gelb1974}
\bibinfo{author}{A.~Gelb}, \bibinfo{author}{J.~F. Kasper},
  \bibinfo{author}{R.~A. Nash}, \bibinfo{author}{C.~F. Price},
  \bibinfo{author}{A.~A. Sutherland}, \bibinfo{title}{{Applied Optimal
  Estimation}}, \bibinfo{publisher}{The MIT Press}, ISBN
  \bibinfo{isbn}{9780262570480}, \bibinfo{year}{1974}.

\end{thebibliography}





\newpage
\section*{Symbols}
\begin{tabular}{l l}
$A_c$ 		& 	Cell surface area, $m^2$						\\
$A_s$ 		& 	Electrode surface area, $m^2$					\\
$a_s$  		& 	Electrode specific interfacial area, $m^{-1}$		\\
$b$ 			& 	Bruggeman coefficient						\\
$c_e$ 		& 	Electrolyte concentration, $mol.m^{-3}$			\\
$c_p$ 		&  	Cell lumped specific heat, $J.kg^{-1}.K^{-1}$ 		\\
$c_s$ 		& 	Solid-phase concentration, $mol.m^{-3}$ 			\\
$c_s^{max}$ 	& 	Active material max concentration, $mol.m^{-3}$ 	\\
$c_s^{surf}$ 	& 	Solid-phase surface concentration, $mol.m^{-3}$ 	\\
$D_e$ 		&	Electrolyte diffusivity, $m^2.s^{-1}$				\\
$D_s$ 		& 	Solid-phase diffusivity, $m^2.s^{-1}$				\\
$E_a^{\psi}$ 	& 	Activation energy of parameter $\psi$, $kJ.mol^{-1}$ \\
$\mathcal{F}$	&	Faraday's constant, $C.mol^{-1}$ 				\\
$h$			& 	Convective heat transfer coefficient, $W.m^{-2}.K^{-1}$	\\
$I$ 			&  	Current, $A$								\\
$i_0$ 		& 	Exchange current density, $A.m^{-2}$			\\
$i_{app}$ 		& 	Applied current density, $A.m^{-2}$				\\ 
$i_e$ 		& 	Current density in electrolyte, $A.m^{-2}$			\\
$i_s$ 		& 	Current density in solid-phase, $A.m^{-2}$		\\
$j^{Li}$ 		& 	Volumetric reaction rate, $A.m^{-3}$				\\
$k$ 			& 	Reaction rate constant, $m^{2.5}.mol^{-0.5}.s^{-1}$	\\
$\dot{q}$ 		& 	Heat generation rate per unit volume, $W.m^{-3}$ 	\\
$R$ 			& 	Gas constant, $J.mol^{-1}.K^{-1}$				\\
$R_c$ 		& 	Contact resistance, $\Omega.m^{2}$ 			\\
$R_s$ 		& 	Radius of solid-phase particles, $m$ 			\\
$T$			& 	Temperature, $K$ 							\\
$T^{ref}$		& 	Standard state reference temperature, $K$ 		\\
$T_\infty$		& 	Coolant temperature, $K$ 					\\
$t_0^+$ 		& 	Li-ions transference number					\\
$U$ 			& 	Electrode open-circuit potential, $V$ 				\\
$V$ 			& 	Cell voltage, $V$ 							\\
$V_c$ 		& 	Cell volume, $m^{3}$ 						\\
\end{tabular}

\subsection*{Greek Symbols}
\begin{tabular}{l l}
$\alpha_a$ 	& 	Anodic charge transfer coefficient			\\
$\alpha_c$ 	& 	Cathodic charge transfer coefficient			\\
$\delta$ 		& 	Thickness of cell layers, $m$ 	\\
$\epsilon_e$ 	& 	Electrolyte volume fraction 				\\
$\epsilon_f$ 	& 	Inert filler volume fraction					\\	
$\epsilon_s$ 	& 	Solid-phase volume fraction 				\\
$\eta$		& 	Overpotential, $V$						\\
$\kappa$ 		& 	Electrolyte ionic conductivity, $S.m^{-1}$		\\
$\theta_s$		& 	Surface solid-phase stoichiometry 			\\
$\theta_s^{avg}$& 	Average solid-phase stoichiometry 			\\
$\theta^0$		& 	Initial solid-phase stoichiometry 			\\
$\rho$ 		& 	Cell bulk density, $kg.m^{-3}$				\\
$\sigma$ 		& 	Solid-phase conductivity, $S.m^{-1}$			\\
$\phi_e$ 		& 	Electrolyte electric potential, $V$			\\
$\phi_s$ 		& 	Solid-phase surface electric potential, $V$		\\
\end{tabular}

\newpage
\listoftables

\newpage
\listoffigures

\end{document}